\documentclass[journal, a4paper]{IEEEtran}
\usepackage[T1]{fontenc}
\usepackage[latin9]{inputenc}
\usepackage{color}
\usepackage{amsmath}
\usepackage{amsthm}
\usepackage{amssymb}
\usepackage{esint}
\usepackage{verbatim}
\usepackage{graphicx}
\usepackage{stfloats}
\usepackage{subfigure}
\usepackage{dsfont}
\usepackage{bm}
\usepackage{multicol}
\usepackage{multirow}
\usepackage{tabularray}
\usepackage{makecell}
\usepackage{algorithm}
\usepackage{algorithmic}
\usepackage{mathtools}
\usepackage{array}
\usepackage[numbers,sort&compress]{natbib}

\makeatletter

\usepackage[normalem]{ulem}

\newtheorem{defn}{Definition}

\definecolor{sblue}{RGB}{0,0,0}
\newcommand\re[1]{\textcolor{sblue}{#1}}
\begin{document}
	\title{ 
		{ LightCom: A  Generative AI-Augmented Framework for QoE-Oriented   Communications
		}
	}
		\author{Chunmei Xu,~\IEEEmembership{Member,~IEEE},  Siqi Zhang, Yi Ma,~\IEEEmembership{Senior Member,~IEEE}, Rahim Tafazolli,~\IEEEmembership{Fellow,~IEEE} 
			
			
			\thanks{C.~Xu, S. Zhang, Y. Ma and, R. Tafazolli are with 5GIC \& 6GIC,
				Institute for Communication Systems (ICS), University of Surrey, Guildford,
				U.K. (emails:\{chunmei.xu; s.zhang; y.ma; r.tafazolli\}@surrey.ac.uk). 
			}
			

		}

		\maketitle
		\begin{abstract}Data-intensive and immersive applications, such as virtual reality, impose stringent quality of experience (QoE) requirements that challenge  traditional  quality of service (QoS)-driven communication systems.  
		This paper presents LightCom, a lightweight encoding and generative AI (GenAI)-augmented decoding framework, designed for QoE-oriented communications under low signal-to-noise ratio (SNR) conditions. 
		LightCom simplifies transmitter design by applying basic low-pass filtering for source coding and minimal channel coding, significantly  reducing processing complexity and energy consumption.
		At the receiver, GenAI models reconstruct high-fidelity content from highly compressed and degraded signals by leveraging generative priors to infer semantic and structural information beyond traditional decoding capabilities. 
		The key design principles are analyzed, along with the sufficiency and error-resilience of the source representation. 
		We also develop importance-aware power allocation strategies to enhance QoE and extend perceived coverage.
		Simulation results demonstrate that LightCom achieves up to a $14$ dB improvement in robustness and a $9$ dB gain in perceived coverage, outperforming traditional QoS-driven systems relying on sophisticated source and channel coding. 
	 	This paradigm shift moves communication systems towards human-centric QoE metrics rather than bit-level fidelity, paving the way for more efficient and resilient wireless networks.
		
		\end{abstract}
		

		\section{Introduction}	
		The rapid growth of data-intensive and immersive applications, such as augmented reality (AR) and virtual reality (VR), is driving new demands in wireless communications \cite{giordani2020toward, wang2023road}. These emerging applications shift the objectives of system design from optimizing traditional quality of service (QoS)---typically quantified by metrics such as data rate, latency, and reliability---toward enhancing the quality of experience (QoE) as perceived by end users \cite{QoE2010, zhang2018towards, zhang2023qoe1, zhao2016qoe, zhang2023qoe2}. Achieving high QoE often necessitates the reliable delivery of semantically meaningful information under stringent resource constraints \cite{shi2021semantic, yang2022semantic}. 
		
\begin{figure*}[tp]
			\vspace{1em}
			\centering
			\includegraphics[width=0.9\textwidth]{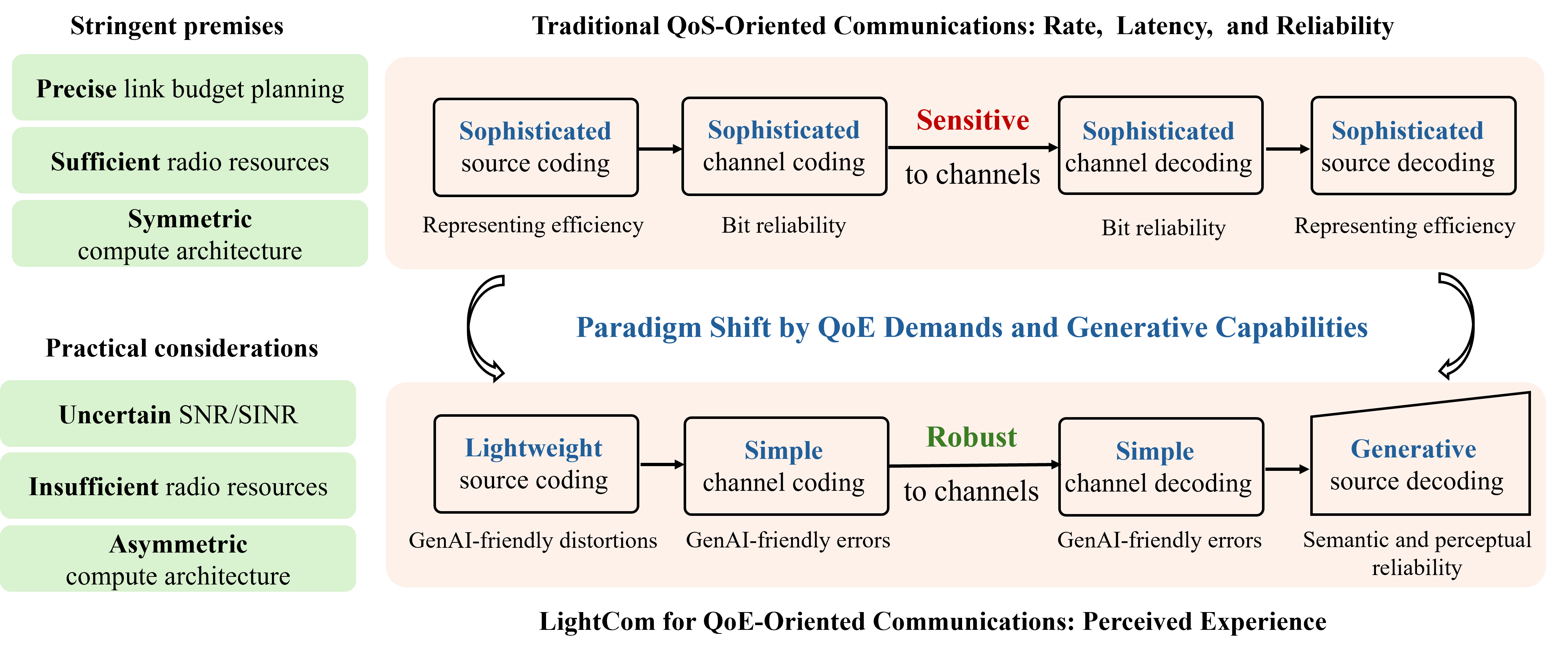}
			
			\caption{{Paradigm shift from QoS-oriented to QoE-oriented communications driven by QoE demands and generative capabilities of GenAI models.}}
			\label{fig:conceptual}
		\end{figure*}

		 This shift in design objectives  introduces significant challenges to traditional communication architectures,  necessitating a fundamental rethinking of system design principles
		 as illustrated in Fig. \ref{fig:conceptual}. Traditional QoS-driven communication systems \re{aim for bit-level fidelity and} depend heavily on sophisticated source and channel coding designs.  	Conventional source coding techniques (e.g., JPEG for images, H.264 for video) aim to maximize representing efficiency while minimizing distortion \cite{wallace1992jpeg, wiegand2003overview}. These schemes exploit inherent data characteristics through transforms to remove redundancy, followed by quantization and entropy coding, resulting in  compact  representations with high inter-symbol dependencies. While such compactness improves representing efficiencies,  it also increases sensitivity to channel-induced errors, making reliable reconstruction heavily dependent on  bit preservation.
 		To address this, traditional channel coding schemes have been intricately developed and advanced to maximize error correction capability, while balancing transmission rate and implementation complexity, ensuring strict bit-level fidelity across varying channel conditions \cite{hamming1950error, viterbi1971convolutional, berrou1993near, richardson2001design}.

	 While theoretically sound, these designs are built upon several strong assumptions \cite{anwar2014rescue, yarkoni2007link, simpulse}. Specifically, \re{they assume predictable radio conditions and stable infrastructure for precise link-budget planning to ensure reliability. Accurate and real-time signal-to-noise ratios (SNR) are required for effective modulation, coding, and link adaptation to approach channel capacity. Additionally, they also assume the symmetric availability of sufficient radio, compute, and energy resources in both transmitter and receiver sides, as the receiver is typically expected to invert the operations of the transmitter.}
	 However, in practical deployments, particularly those involving energy- and compute-constrained transmitters such as IoT devices \cite{khanh2022wireless}, wearables, or mobile nodes, and in unpredictable environments like post-disaster areas (e.g., during earthquakes or tsunamis) \cite{khanh2022wireless,wang2023overview}, these assumptions are increasingly difficult to hold.  Furthermore, the conventional emphasis on bit-level fidelity, while optimal from an information-theoretic perspective, does not necessarily correspond to high QoE in tasks involving semantic or perceptual interpretation \cite{gunduz2022beyond, xu2025generative}.  For instance, a user may prefer a semantically coherent but lossy image over a perfectly transmitted yet incomplete one \cite{liu2025resitok}. These observations highlight the vulnerability of  bit-centric and symmetric designs under real-world constraints, and motivate the development of alternative architectures that gracefully degrade signals,  prioritize meaning over exact bits, and exploit computational asymmetry at both ends.

		Recent advances in generative artificial intelligence (GenAI), including diffusion models and transformer-based decoders, offer a compelling alternative for QoE-oriented communication system design \cite{chen2024big,10384630}. 
		Trained on large-scale, multimodal datasets spanning text, audio, images, and video,  these  models, comprising billions of neural networks (NN) parameters, exhibit strong generative priors and are capable of  inferring and reconstructing high-fidelity content from sparse, or degraded inputs \cite{radford2021learning, brown2020language, rombach2022high,zhang2023adding, yu2024scaling}.  
		When deployed at the receiver, GenAI model enables semantic inference of the transmitted message and reconstruction of perceptually satisfactory content even under poor channel conditions.
		Consequently, compression distortions and residual decoding errors can be compensated by GenAI-augmented receiver, effectively decoupling perceptual reliability from  strict bit-level accuracy. 
		This enables a new design paradigm in which the transmitter does not necessarily  deliver every bit faithfully, so long as the receiver can infer the intended meaning from the received signal with satisfactory QoE.

		Motivated by this insight, we propose {\it{LightCom}},  a QoE-oriented communication framework, that leverages computational asymmetry between the transmitter and receiver.  LightCom employs an ultra-lightweight transmitter that applies lightweight source encoding and simple forward error correction codes (FEC), thereby drastically reducing computational burdens and energy consumption.  In contrast,  the receiver is augmented with powerful GenAI models  capable of reconstructing semantically meaningful and perceptual satisfactory content from distorted and degraded representation received.    This asymmetric design paradigm takes advantage of the increasingly abundant computational resources at the receiver side, such as base stations and edge servers, to compensate for the transmitter's limited transmission capability. 
	 	The rationale behind LightCom is threefold.  First, the lightweight source encoding reduces inter-symbol dependencies trading off representation efficiency while preserving QoE-critical components, in order to produce representations that are well-suited for GenAI models to reconstruct QoE-satisfactory content. Second, the use of simple channel codes mitigates error propagation and prevents burst errors,  resulting in  more independent and less structured   decoding errors that can be semantically and perceptually corrected by the GenAI model through generative inference. Third, the GenAI model at the receiver enables high-fidelity reconstruction by leveraging strong generative priors captured from data training, effectively shifting robustness from physical-layer reliability to high-level semantic inference.  

		Importantly, LightCom departs from existing joint source-channel coding (JSCC) approaches in semantic communications using deep learning techniques \cite{bourtsoulatze2019deep, weng2021semantic, erdemir2023generative, tong2024multimodal}.  Prior works in deep JSCC typically rely on end-to-end NN architectures, which require substantial and symmetric computational resources on both ends. Moreover, these models are often tailored to specificdata modalities and channel conditions, limiting their generalization and practicality in real-world deployments \cite{xu2024semantic, qiao2024latency}.  Besides, deep JSCC approaches \re{lack} underlying degree of freedom for optimization and face compatibility challenges with existing digital infrastructures. Recent studies have shown that separate source and channel coding, with the deployment large language models, achieved superior performance than deep JSCC \cite{ren2025separate}. In contrast, LightCom separates the roles of source and channel coding: the transmitter performs minimal  processing, while the receiver reconstructs  contents using a general-purpose GenAI model. The resulting asymmetry in compute capabilities between transmitter and receiver provides improved energy efficiency, enhanced robustness, and greater adaptability across diverse communication environments.

		The main contributions of this work are summarized as follows:
	 
		\begin{itemize}
			\item We propose {LightCom}, a novel asymmetric computing framework for QoE-oriented communications, featuring a lightweight transmitter and a GenAI-augmented receiver. The  transmitter applies basic low-pass filtering for source coding and minimal channel encoding when necessary, while the receiver employs a pre-trained GenAI model  to semantically and perceptually reconstruct content by leveraging strong generative priors.
			
			\item We characterize the sufficiency and error-resilience  of the source representation within the LightCom framework, and analyze two key design principles: \re{low-pass filtering (LPF)} for semantically source compression, and weak channel coding (WCC) for robustness and compatibility with generative reconstruction.
			 
			\item We develop a  hybrid QoE metric that combines two complementary QoE-driven measures: the natural image quality evaluator (NIQE) for perceptual quality assessment, and contrastive language-image pretraining (CLIP)-based similarity for semantic coherence evaluation. 
			Based on this metric, the perceived coverage is defined and quantified as the region within which the QoE requirement is guaranteed, revealing joint  impact of   compression rate and channel conditions. 
	 			
			\item We design importance-aware power allocation strategies to improve QoE and extend perceived coverage. Waterfilling (WF)-based algorithms are developed  for both channel-uncoded and channel-coded cases, enabling adaptive and efficient resource allocation based on data importance and channel conditions.
	
		\end{itemize}
		
		The proposed LightCom framework is validated through extensive simulations using image data. 	Results show that LightCom achieves greater robustness than traditional QoS-oriented communication systems in low-SNR scenarios,  with SNR reductions of up to $14$ dB and $4.5$ dB at a compression of $9\%$ in channel-uncoded and channel-coded cases, respectively.   
		Moreover, it provides substantial perceived coverage extension, achieving  $7.5-8$ dB gains at near-lossless compression with rates ranging from $r=33\%$ to $50\%$.  Notably, these improvements stem not from enhanced bit-level reliability, but rather from the receiver's strong ability to reconstruct perceptually and semantically meaningful content via generative inference. These results highlight the potential of the proposed LightCom framework  to enhance robustness, extend coverage, and reduce retransmissions in QoE-oriented communications,  particularly under uncertain SNR/SINR conditions and constrained resource scenarios.


			\section{LightCom Framework \label{sec:II}}
\begin{figure}[t!]
				\centering
				\includegraphics[width=0.45\textwidth]{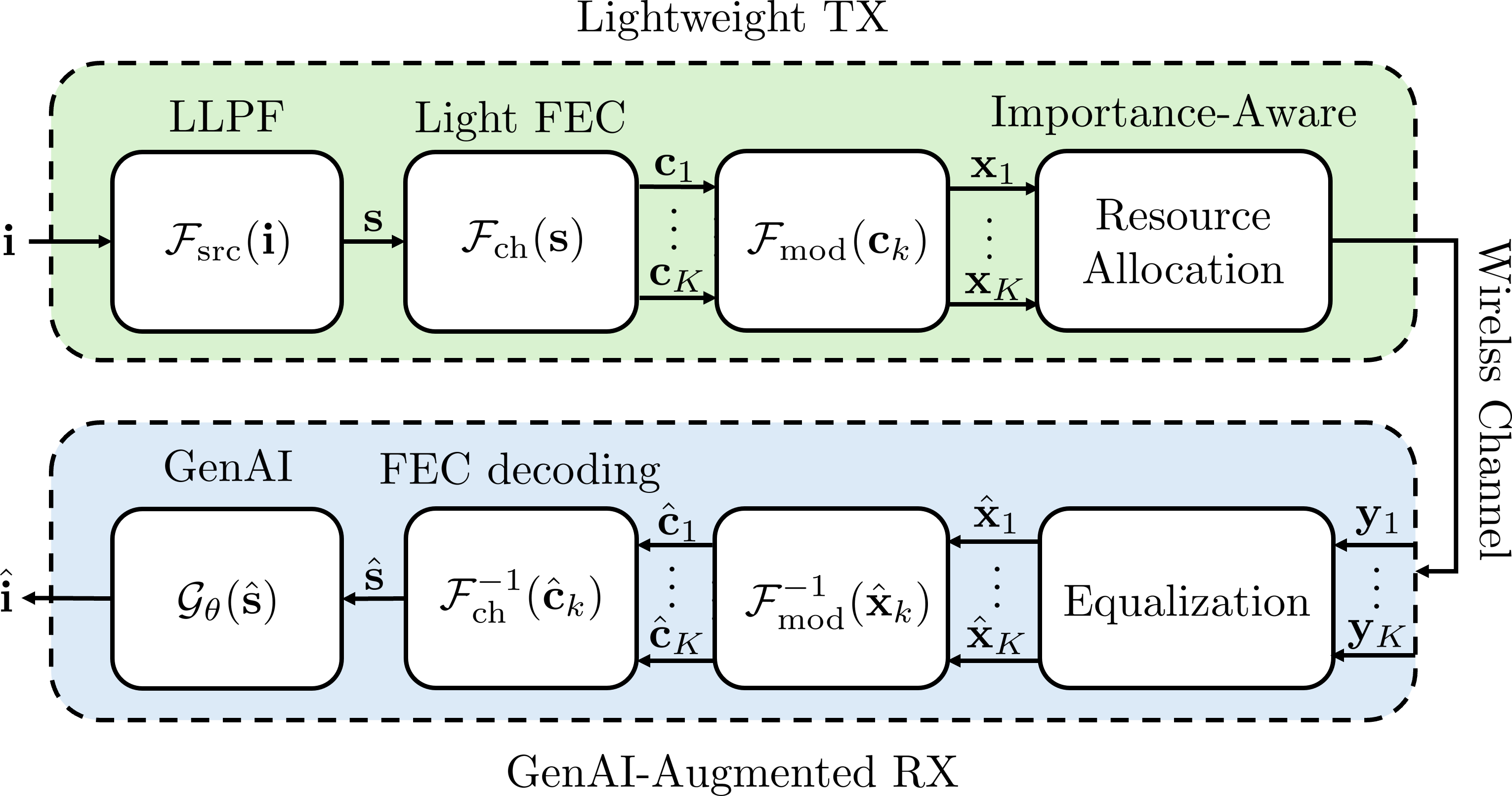} 
				\caption{Schematic of LightCom transmitter and receiver architecture.}\label{lightcom}
			\end{figure}

			The LightCom framework, depicted in Fig. \ref{lightcom}, is founded on the principle of computational asymmetry, wherein a lightweight transmitter is paired with a  GenAI-augmented receiver to enable the reconstruction of semantically meaningful content under resource-constrained and unreliable channel conditions. This section formalizes the system architecture, key signal processing blocks, and core design motivations.
			
			LightCom can be understood as a split-inference generative communication system. 
			The transmitter performs minimal pre-processing, akin to a truncated front-end of a large generative model, while the receiver hosts a full generative decoder. 
			This architecture draws parallels to edge-cloud AI pipelines \cite{banitalebi2021auto}, where partial model computation is performed at the edge and offloaded to powerful back-end servers. However, unlike digital systems with error-free links, LightCom operates over a noisy wireless channel with bandwidth and reliability constraints. This architectural separation allows the transmitter to significantly reduce energy consumption and complexity by transmitting only a coarse, lossy representation of the original signal. The receiver, equipped with pre-trained semantic priors, reconstructs high-fidelity outputs, effectively shifting robustness from physical-layer reliability to high-level semantic inference.

			\subsection{Lightweight Transmitter}
			For QoE-oriented communication,  we observe that fine-grained details (e.g. high-frequency components in image/video data) often contribute less to semantic understanding but perceptual quality. Thereby, these components can be selectively distorted at the transmitter to remain semantic similarity, and regenerated at the receiver using powerful GenAI models to improve perceptual quality. This observation aligns with Shannon's information theory framework, where {\re high-probability events}, corresponding to high-frequency details, carry less self-information (quantified as $-\log p_i$). On this basis, we propose a lightweight source coding scheme based on  \re{ low-pass filtering (LPF)} that preserves essential low-frequency components while discarding high-frequency ones.

			Let \( \mathbf{i} \in \mathbb{R}^I \) denote the original input signal (e.g., a flattened image or sensor measurement). 
			The transmitter applies a lightweight transformation:
			\begin{equation} \label{eq01}
				\mathbf{s} = \mathcal{F}_{\mathrm{src}}(\mathbf{i}),
			\end{equation}
			where \( \mathcal{F}_{\mathrm{src}} \) denotes the source encoder such as \re{LPF}, and \( \mathbf{s} \in \mathbb{R}^S \), with \( S \ll I \). Specifically, for an input measurement $\mathbf{I}$ of size $H\times W$ (flattened into $\mathbf{i}$ of size $I=HW$), $\mathbf{I}$ is divided into non-overlapping blocks of uniform size  $B_1\times B_2$. Applying low-pass filter in the $(i,j)$-th block produce  the $(i,j)$-th element of the compressed representation $\mathbf{S}$ (flattened into $\mathbf{s}$ of size $S=rI$ with $r=1/{B_1B_2}$ the compression rate.) The computational complexity of this approach is minimal, which requires only $1$ multiplication and $B_1B_2$ additions per block for mean filter.

			The pixel elements of $\mathbf{s}$ are partitioned according to sub-pixel importance (i.e., the bit position  \cite{xu2025dataimportanceJ}), creating $K$ bit sequences $\mathbf{s}_k$ of equal length. Apply the lightweight channel encoding:
			\begin{equation}\label{eq02}
				[\mathbf{c}_1, \dots, \mathbf{c}_K]= \mathcal{F}_{\mathrm{ch}}(\mathbf{s}_1, \dots, \mathbf{s}_K),
			\end{equation}
			where  \( \mathcal{F}_{\mathrm{ch}} \) denotes a minimal channel encoding function employing weak channel codes (WCC), which is applied to these bit sequences. The WCC may involve no coding, simple repetition, spreading, or basic parity checks, yielding $K$ channel-coded outputs \( \mathbf{c}_k \in \mathbb{R}^S \).

			These coded bit sequences are then modulated:
			\begin{equation}\label{eq02}
				[\mathbf{x}_1, \dots, \mathbf{x}_K]= \mathcal{F}_{\mathrm{mod}}(\mathbf{c}_1, \dots, \mathbf{c}_K),
			\end{equation}where $\mathcal{F}_{\mathrm{mod}}$ denotes a modulation function, generating $K$ sub-streams $\mathbf{x}_k$ with equal length of $L$. 
			For the $k$-th sub-stream, the transmitted signal $\mathbf{x}_k$ is normalized such that $\mathbb E[\mathbf{x}_k\mathbf{x}_k^\mathrm{H}]=\mathbf{I}_{L\times L}$, where $\mathbb{E}(\cdot)$ stands for the expectation and $(\cdot)^\mathrm{H}$ for the Hermitian.  
			
			These sub-streams are transmitted over orthogonal sub-channels, with received signal for the $k$-th sub-stream expressed as:
			\begin{equation}
				\mathbf{y}_{k} = h_k\sqrt{p_k}\mathbf{x}_k + \mathbf{n}_k,
			\end{equation}where $h_k$ denotes the sub-channel coefficient. $p_k$ is the allocated power per symbol in the sub-stream, and $\mathbf{n}_k$ is Gaussian noise with each entry following $\mathcal{CN}(0,\sigma^2)$. 
			
			\subsection{GenAI-Augmented Receiver}
			The receiver performs standard physical-layer operations, including channel equalization and demodulation and channel-decoding to recover sub-streams $\hat{\mathbf{s}}_{k}$, which are then aggregated to reconstruct  $\hat{\mathbf{s}}$. Assuming perfect channel state information (CSI) and precise power budget known to the receiver, The received SNR of the $k$-th sub-stream is:
			\begin{equation}
				\mathrm{snr}_k = \frac{p_k\vert h_k \vert^2}{\sigma^2}. 
			\end{equation}
			Note that $\mathrm{snr}_k$ becomes uncertain under channel dynamics and impairments, or unavoidable interference in practical scenarios.  The decoded signal is:
			\begin{equation} \label{eq04}
				\hat{\mathbf{s}}_k = \mathcal{F}_{\mathrm{ch}}^{-1}(\mathcal{F}_{\mathrm{mod}}^{-1}(\hat{\mathbf{x}}_k)),
			\end{equation}where $\hat{\mathbf{x}}_k=(h_k\sqrt{p_k})^{-1}\mathbf{y}_k$ is the soft message of ${\mathbf{x}}_k$.
			
			Instead of applying a classical source decoder, the receiver uses a generative model \( \mathcal{G}_\theta \) to reconstruct the source:
			\begin{equation}\label{eq05}
				\hat{\mathbf{i}} = \mathcal{G}_\theta(\hat{\mathbf{s}}),
			\end{equation}
			where \( \mathcal{G}_\theta: \mathbb{R}^S \rightarrow \mathbb{R}^I \) is a pre-trained GenAI model such as a diffusion model or transformer-based decoder. Trained on large datasets, it captures semantic priors and can infer plausible reconstructions even when \( {\mathbf{s}} \) is distorted and \( \hat{\mathbf{s}} \) is degraded with residual decoding errors.
			
			This shift enables the receiver to recover meaning and perceptual quality from imperfect transmissions, making the system robust under uncertain SNRs, low link budgets, and asymmetric compute scenarios.

			\subsection{{ Residual Decoding Errors}\label{sec3}}

			
			Due to noisy nature of wireless channels, errors may occur in the recovered sub-streams $\hat{\mathbf{s}}_{k}$. For WCC, we consider both uncoded transmission and scenarios employing basic parity-check codes. 
			
			In the uncoded case, the bit error rate (BER) of the $k$-th sub-stream, denoted as $\mathrm{ber}_k^\mathrm{u}$, under $M$-QAM modulation is given by \cite{goldsmith2005wireless}:
			\begin{equation}\label{eq:BER_SNR_Uncoded}
				\mathrm{ber}_k^\mathrm{u} = \alpha_\mathrm{u} \mathcal{Q}\left(\beta_\mathrm{u}\sqrt{\mathrm{snr}_k}\right),
			\end{equation}where $\alpha_\mathrm{u} =  \frac{4}{\log_2 M}(1-\frac{1}{\sqrt{M}})$,  $\beta_\mathrm{u}=\sqrt{\frac{3}{M-1}}$, and $\mathcal{Q}(x)=\frac{1}{\sqrt{2\pi}}\int_{x}^{\infty}\mathrm{exp}(\frac{-t^2}{2})dt$ is the Q-function. Since SNR is always non-negative, $\mathrm{ber}_k^\mathrm{u}\le \frac{1}{2}\alpha_\mathrm{u}$. 
			
			In the channel-coded case, the BER  of the $k$-th sub-stream, denoted as $\mathrm{ber}_{k}^\mathrm{c}$, can be approximately modeled as:
			\begin{equation}\label{eq:BER_SNR_Coded}
				\mathrm{ber}^\mathrm{c}_k \approx \alpha_\mathrm{c}\exp\left(\beta_\mathrm{c}\mathrm{snr}_k\right).
			\end{equation}Here, $\alpha_\mathrm{c}\ge 0$ and $\beta_\mathrm{c}\le 0$ are parameters determined by the specified channel coding and modulation schemes, which can be obtained through data fitting \cite{xu2025dataimportanceJ,xu2025dataimportance}.  
			
			These bit errors lead to erroneous compressed source $\mathbf{s}$, specifically affecting its pixel units $\mathbf{s}_{ij}$.  	With the assumption of equal importance among pixels, the degradation can be quantified by the importance-weighted mean square error (IMSE) \cite{xu2025dataimportanceJ}:
			\begin{align}\label{eq:MSE_BER}
				\mathrm{IMSE}(\hat{\mathbf{s}}, \mathbf{s}) &= \sum_{k}  \gamma_k \frac{\Vert \hat{\mathbf{s}}_{k}-\mathbf{s}_{k}\Vert^2 }{S}\nonumber \\ 
				&  \approx \gamma_k\mathrm{ber}_k,
			\end{align}where  $\gamma_k=2^{2(k-1)}$ is the importance weight of   $\mathbf{s}_k$, and
			$\mathrm{ber}_k\in \{\mathrm{ber}_k^\mathrm{u}, \mathrm{ber}_k^\mathrm{c}\}$ is the BER of the $k$-th sub-stream. Note that \eqref{eq:MSE_BER} is an approximation of the MSE between $\hat{\mathbf{s}}$ and $\mathbf{s}$, by assuming  at most one bit per pixel is incorrectly recovered.

 		\section{Principles of LightCom}		
 		 LightCom is designed to prioritize perceptual and semantic
 		 quality, rather than traditional bit-level fidelity. This shift
 		 demands a rethinking of the communication architecture,
 		 especially the interplay between source encoding, channel
 		 robustness, and generative reconstruction.  In this section, we first examine  the sufficiency and error-resilience of the compressed representation $\mathbf{s}$, and analyze the key design principles of \re{LPF}  for semantically aware source compression, and the use of weak channel codes (WCCs) to enhance robustness and compatibility with generative reconstruction. Without loss of generality, it is assumed that a lower value of QoE performance metric (detailed in Sec. \ref{sec:III}) corresponds to better system performance.

 		\subsection{Sufficiency and Error-Resilience}
 		
 		A source representation $\mathbf{s}$ is sufficient when the QoE requirement is guaranteed.  To characterize this sufficiency, we first give the definition of
 		QoE-essential representation $\mathbf{s}^*$, which is given as follows:
			\begin{defn}
				The QoE-essential representation $\mathbf{s}^*$ is both sufficient to guarantee the QoE requirement, and necessary in that any distortion leads to a unacceptable QoE, which can be expressed as:
				{\begin{equation}
						\mathbf{s}^* \in \mathcal S^* \triangleq \left\{ 
						\mathbf{s} \,\middle|\,
						\begin{aligned}
							& \mathrm{QoE}(\hat{\mathbf{I}}_{\mathbf{s}};\mathbf{I}) \le \mathrm{QoE}_\mathrm{th} \\
							& \mathrm{QoE}(\hat{\mathbf{I}}_{\tilde{\mathbf{s}}}; \mathbf{I}) > \mathrm{QoE}_\mathrm{th} 
						\end{aligned}
						\right\},
				\end{equation}}where $\tilde{\mathbf{s}}$ is any distorted version of $\mathbf{s}$,
			 $\hat{\mathbf{I}}_{\mathbf{s}}=\mathcal{G}_\theta(\mathbf{s})$, and $\hat{\mathbf{I}}_{\tilde{\mathbf{s}}}=\mathcal{G}_\theta(\tilde{\mathbf{s}})$. Here,  $\mathrm{QoE}(\cdot; \cdot)$ is the QoE function, and $\mathrm{QoE}_\mathrm{th}$ denotes the QoE requirement.  
			\end{defn}

\begin{figure}[tp]
				\centering
				\includegraphics[width=0.9\columnwidth]{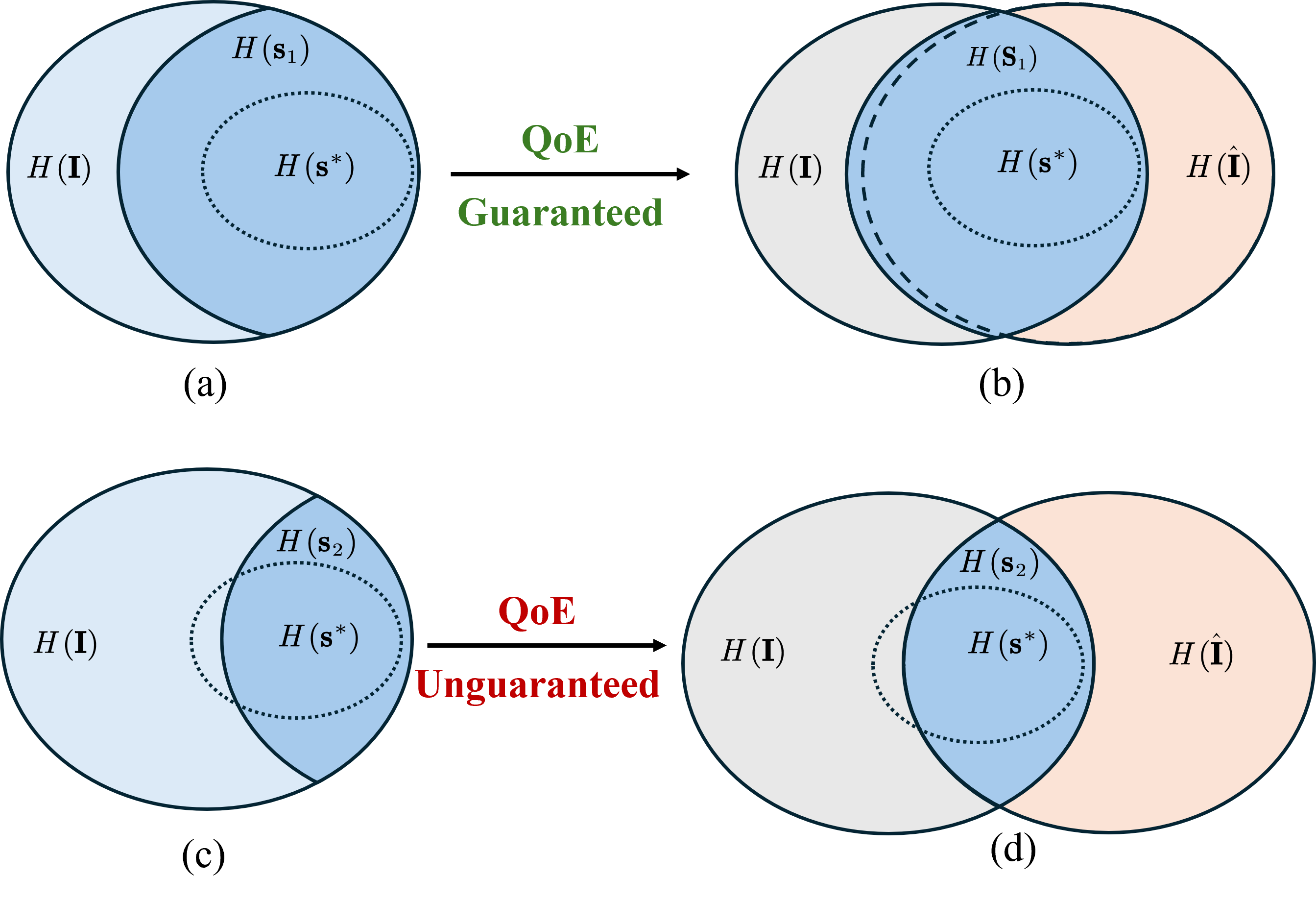}
				\caption{ Diagram of sufficient and insufficient representations, where orange regions represent the generated content by the GenAI-augmented receiver.}
				\label{fig:entropy}
			\end{figure}

			Fig. \ref{fig:entropy} illustrates the QoE performance of the regenerated source under different representations, highlighting the characteristic of sufficiency and the role of $\mathbf{s}^*$. The orange regions of $\hat{\mathbf{I}}$ 
			denotes semantic information inferred by the GenAI-augmented receiver, which is generally less critical. The QoE requirement is guaranteed when  $\mathbf{s}^*$ is preserved in the reconstructed content $\hat{\mathbf{I}}$, as demonstrated in Figs. \ref{fig:entropy}(b). 
			It is important to note that $\mathbf{s}_1$ may undergo some distortion during the regeneration process, consistent with the data processing inequality.  Conversely, as shown in Fig. \ref{fig:entropy}(d), $\mathbf{s}^*$ is not preserved, the QoE requirement cannot be guaranteed.

			Next, we explore the error-resilience of the GenAI-augmented receiver, reflecting the impact of residual channel errors on generative reconstruction. 
			To contextualize this, we compare it with  compact representations and strong error protection used in traditional communication systems. 
			While such compactness enhances compression efficiency, it also introduces strong inter-symbol dependencies, making source decoding highly sensitive to even minor residual channel errors. Consequently, strong channel codes, such as LDPC, are typically employed to ensure bit-level reliability.
			In contrast,  error-resilient representations consist of more independent symbols, sacrificing compression efficiency in favor of robustness, making GenAI-augmented receiver  more tolerate to isotropic and unstructured residual channel errors. 
			Even when some symbols are corrupted, correctly decoded symbols still retain meaningful information that contributes to the generative reconstruction process. The definition is given as follows:

		\begin{defn}\label{def2}{Error-resilience} refers to the capability of GenAI-augmented receiver to reconstruct  the source content from a corrupted representation $\hat{\mathbf{s}}$ such that
					\begin{equation}\mathbb E \big[\mathrm{QoE}(\hat{\mathbf{I}}_{\hat{\mathbf{s}}};\mathbf{I})\big] \le \mathrm{QoE}_\mathrm{th},
				\end{equation}where:
				 \begin{equation}
				 	\mathrm{IMSE}(\hat{\mathbf{s}}, \mathbf{s}) \le \mathrm{IMSE}_{\max}.
				\end{equation}$\mathrm{IMSE}_{\max}$ denotes the maximum tolerable  channel error, and characterizes how resilience the LightCom is to the channels. It is jointly determined by the sufficient source representation $\mathbf{s}$ and the residual error types.
			\end{defn}
			
			The QoE performance of $\hat{\mathbf{I}}_{\hat{\mathbf{s}}}$ generally degrades with increasing IMSE, thereby motivating adaptive radio resource allocation to enhance QoE by minimization IMSE. This is  particularly important given the lack of an explicit analytical expression linking QoE to SNR. 


				\subsection{Design Principles}
				Based on the above analysis, we now discuss the key design principles of \re{LPF} for source coding and WCC for channel coding in support of robust  QoE-oriented communications. 
				\subsubsection{\re{LPF} for source coding} For image and video applications, the lightweight \re{LPF} source encoder produces a lossy representation by filtering out high-frequency details. This results in a low-quality, blurred version of the original image, with significantly reduced inter-symbol dependencies compared to traditionally sophisticated source coding schemes. 
				
				The process of recovering high-quality outputs from such degraded inputs is known in the computer vision community as image restoration (IR). The goal of IR is to reconstruct high-quality images from degraded inputs suffering from low resolution, blur, or noise. Recent advances have achieved remarkable results by leveraging powerful pre-trained generative models \cite{lin2024diffbir, yu2024scaling}. These models learn strong generative priors that capture the inherent structure of natural images, enabling restoration outputs that align well with the natural image distribution. In particular, \cite{yu2024scaling} introduced the pioneering Scaling-UP IR (SUPIR) method, which demonstrated exceptional capabilities in ultra-high-quality image reconstruction.  The pre-trained SUPIR model can be deployed at compute-rich receivers to augment reconstruction performance, which leverages asymmetric compute capabilities. The sufficiency of representation is determined by the size of \re{LPF}, or equivalently, the compression rate.  

				\subsubsection{WCC for error protection} Given the \re{LPF} source encoder, the resilience of LightCom for QoE-oriented communications  can be characterized by the set of SNRs that guarantee the QoE requirement:	
					\begin{equation}
					\left\{ 
					\mathrm{snr} \,\middle|\, \mathbb E[\mathrm{QoE}(\hat{\mathbf{I}}_{\hat{\mathbf{s}}}; \mathbf{I})] \le \mathrm{QoE}_\mathrm{th} 
					\right\},
				\end{equation}where  $\hat{\mathbf{s}}$ denotes a degraded version of  the transmitted representation $\mathbf{s}$ under a given $\mathrm{snr} $.   Since the MSE is approximately decreasing with SNR,  according to \eqref{eq:BER_SNR_Uncoded},  \eqref{eq:BER_SNR_Coded}, \eqref{eq:MSE_BER}, a larger $\mathrm{MSE}_{\max}$ indicates greater resilience of LightCom to channel dynamics and impairments.
				
				In LightCom, the transmitter employs minimal channel coding with WCC, resulting in a higher likelihood of residual errors at the receiver post-decoding. Denote the channel-decoded representation as:
				\begin{equation}\label{eq17}
					\hat{\mathbf{s}} = \mathbf{s} + \mathbf{e},
				\end{equation}
				where $\mathbf{e}$ denotes the residual error vector introduced by imperfect channel decoding.
			
				To analyze the impact of $\mathbf{e}$, we analyze two representative classes of channel codes: strong channel codes (SCCs) and WCCs. SCCs (e.g., LDPC, Tubor)  typically introduce structured dependencies across symbols, which can lead to bursty or correlated residual errors due to decoder convergence failures or parity-check violations.   In contrast, WCCs, which include schemes like no coding, repetition, or Hamming codes, often lacks the structural correlation mechanism of SCC, limiting error propagation and making errors more independent and unstructured.  We model these two residuals errors as follows:

				 \textbf{Weak-Code Errors:} $\mathbf{e}_w \sim \mathcal{N}(0, \sigma_w^2 \mathbf{I})$, capturing uncorrelated errors from the absence of channel coding or simple schemes such as repetition or Hamming codes.
				 
				 \textbf{Strong-Code Errors:} $\mathbf{e}_s \sim \text{Burst}(\sigma_s^2, L_s)$, modeling long bursty residuals caused by decoding failures in strong codes, where $L_s$ denotes the burst length.
		 
		 		Residual channel errors shift the distribution of the decoded representation $\hat{\mathbf{s}}$ away from the distribution seen during the generative model's training \cite{lin2024diffbir,yu2024scaling}, thereby degrading reconstruction performance. For the same error variance, i.e., $\sigma_w^2 = \sigma_s^2$, spatially unstructured Gaussian-like noise from WCC results in a smaller distributional shift compared to structured, bursty corruption from SCC. Therefore, the GenAI-augmented receiver generally exhibits higher robustness when WCC is used for error protection, indicating that the tolerable $\mathrm{MSE}_{\max}$ under WCC is higher than that under SCC.


		 \section{QoE Metrics and Perceived Coverage\label{sec:III}}
		 
		 \re{This section introduces the QoE metrics used to evaluate the performance of the proposed LightCom framework, which also serve as key indicators for system design. While coverage has long been a central challenge in wireless communications, existing solutions often incur high cost and complexity. In this work, we define and quantify {perceived coverage} for QoE-oriented communications based on the developed QoE metrics, offering a low-cost and QoE-driven perspective on coverage extension.}

 	\subsection{Evaluation Metrics\label{sec3.A}}
	Traditional metrics like bit error rate (BER) and peak signal-to-noise ratio (PSNR) fail to capture perceptual  and semantic fidelity, which are crucial in QoE-oriented systems. To evaluate the QoE performance, we use two complementary metrics based on NIQE and CLIP.
	
	\subsubsection{NIQE-based perceptual fidelity}
 	The developed lightweight source encoder suppresses high-frequency details, resulting in a blurred compressed representation ${\mathbf{s}}$. Transmission over noisy wireless channels further introduces errors, producing a noisy version $\hat{\mathbf{s}}$. Both blurring and noise  disrupt the naturalness of source content, thereby affecting the perceived quality of the reconstructed output. This naturalness can be  effectively assessed using the no-reference NIQE metric based on natural scene statistics \cite{mittal2012making}. 
 	
 	This observation motivates the use of NIQE to measure the perceived quality in terms of naturalness, which is defined as a function of the  compression rate and SNR: 
 	\begin{equation}
 		D_\mathrm{NIQE} (\mathrm{snr}, r) = \mathbb E_{\mathbf I,\hat{\mathbf{I}}}\big[\mathrm{NIQE}(\hat{\mathbf{I}})\big],
 	\end{equation}where:
 	\begin{equation}
 	\mathrm{NIQE}(\hat{\mathbf{I}})= \sqrt{(\boldsymbol{\mu}_{\hat{\mathbf{I}}} - \boldsymbol{\mu}_o)^T \left( \frac{\boldsymbol{\Sigma}_{\hat{\mathbf{I}}} + \boldsymbol{\Sigma}_o}{2} \right)^{-1} (\boldsymbol{\mu}_{\hat{\mathbf{I}}}- \boldsymbol{\mu}_o)},
 	\end{equation}where $\hat{\mathbf{I}}$ is reconstructed under the   compression rate of $r$  and SNR of $\mathrm{snr}$. $\boldsymbol{\mu}_{\hat{\mathbf{I}}}, \boldsymbol{\Sigma}_{\hat{\mathbf{I}}}$ are the sample statistics from ${\hat{\mathbf{I}}}$, and $\boldsymbol{\mu}_o, \boldsymbol{\Sigma}_o$ are from pristine natural images. Consistent with NIQE's scoring range, $D_\mathrm{NIQE} (r, \mathrm{snr})\in [0,100]$, with lower values indicating higher perceptual quality.

 	\subsubsection{CLIP-based semantic distance}
 	However, $D_\mathrm{NIQE}$ alone is  insufficient for comprehensive  evaluation, as a reconstructed source with low $D_\mathrm{NIQE}$  does not necessarily maintain high semantic similarity to the original one. For example, a jammer might cause the receiver to decode an unrelated but visually natural source, indicating high naturalness despite communication failure.  The high-level semantic conveying can be evaluated through semantic consistency, which can be measured using the CLIP score \cite{xu2024semantic}.  
 	We define a semantic distance as a function of compression rate and SNR:
 	\begin{align}
 		D_\mathrm{CLIP}&(\mathrm{snr}, r) = \mathbb{E}_{\mathbf{I},\hat{\mathbf{I}}}\big[ 1- \mathrm{{CLIP}}(\mathbf{I}, \hat{\mathbf{I}})\big],
 	\end{align}where:
 		\begin{equation}\label{eq07}
 		\text{CLIP}(\mathbf{I}, \hat{\mathbf{I}}) = \frac{\phi(\mathbf{I})^T \phi(\hat{\mathbf{I}})}{\|\phi(\mathbf{I})\| \cdot \|\phi(\hat{\mathbf{I}})\|},
 	\end{equation}
 	where \( \phi(\cdot) \) denotes the CLIP encoder (i.e. embedding function) \cite{radford2021learning}, and $(\cdot)^T$ denotes the transpose.  With $\mathrm{CLIP}(\mathbf{I}, \hat{\mathbf{I}}) \in [-1,1]$, the semantic distance falls within $D_\mathrm{CLIP} \in [0,2]$, where lower values indicate higher semantic similarity. 
 	Higher values of CLIP imply stronger semantic preservation.
 	
   	Together, $D_\mathrm{NIQE}$ and $D_\mathrm{CLIP}$ provide an effective framework for assessing LightCom's performance under practical, perception- and meaning-driven communication goals. Therefore,  we propose a QoE metric:
 	 
 	\begin{equation}\label{eq:QoE}
 		\mathrm{QoE}(\mathrm{snr}, r) =   
 			[D_\mathrm{NIQE} (\mathrm{snr}, r), D_\mathrm{CLIP}(\mathrm{snr}, r)].
 	\end{equation}

		 \subsection{Perceived Coverage} 
		Conventional coverage refers to regions where signals can be reliably received. This is typically characterized using Shannon's channel coding theorem, where communication is reliable with an arbitrary low probability of error when the transmission rate is below the channel capacity; otherwise, an outage is declared.  This definition emerged from a technical communication perspective that targeted on accurate bit-level recovery \cite{weaver1953recent}. Such definition becomes inadequate for QoE-oriented communications that focus on subjective experience. Based on the proposed QoE metric, we define \emph{perceived coverage} as the set of compression rate and channel condition tuples, under which the reconstructed signal remains semantically and perceptually acceptable:
			\begin{defn} 
				Perceived coverage refers to the set of compression rate and SNR combinations that guarantee the QoE requirement $\mathrm{QoE}_\mathrm{th}\triangleq [D_\mathrm{NIQE}^\mathrm{th}, D_\mathrm{CLIP}^\mathrm{th}]$.  Mathematically, it is expressed as: 
				 \begin{equation}
					\mathcal C = \left\{ \left(\mathrm{snr}, r\right)\middle|\begin{array}{c}
						D_{\mathrm{NIQE}}(\mathrm{snr}, r)\le D_{\mathrm{NIQE}}^\mathrm{th}\\
						D_{\mathrm{CLIP}}(\mathrm{snr}, r)\le D_{\mathrm{CLIP}}^\mathrm{th}
					\end{array}\right\},
				\end{equation}revealing joint dependence on both compression rate and channel conditions. 
			\end{defn}
			
\begin{figure}[tp]
				\centering
				\includegraphics[width=1\columnwidth]{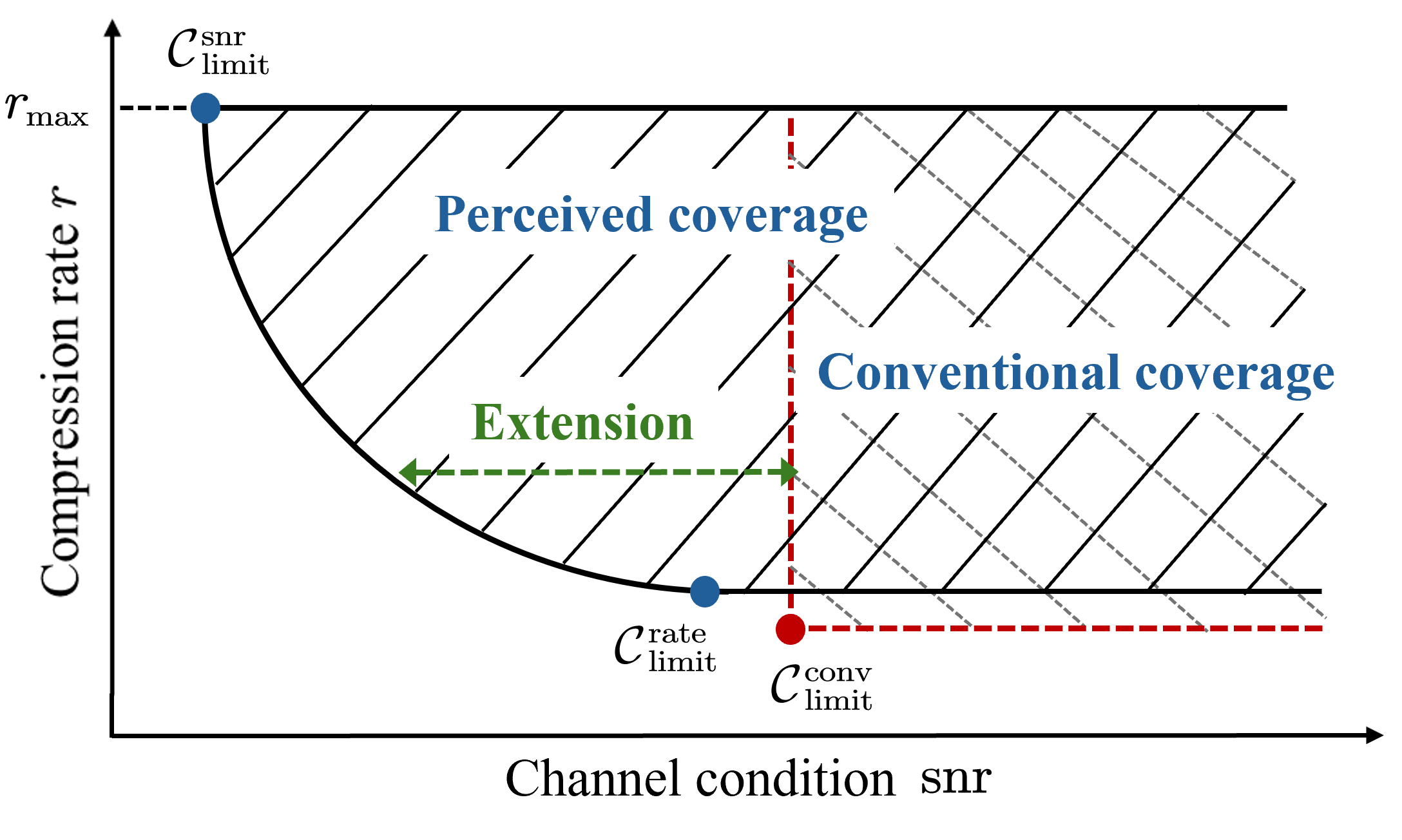}
				\caption{{Perceived coverage and conventional converge under fixed transmission rate, where  $\mathcal C_{\mathrm{limit}}^{\mathrm{snr}}=({\mathrm{snr}}^{\vdash},\underline{r}_{\mathrm{max}})$, $\mathcal C_{\mathrm{limit}}^{\mathrm{rate}}=({\mathrm{snr}}^{\dashv},\underline{r}_{\mathrm{min}})$, and $\mathcal C_{\mathrm{limit}}^{\mathrm{conv}}=({\mathrm{snr}}_{\mathrm{conv}}^{\vdash},\underline{r}_{\mathrm{conv}})$. }}
				\label{fig: coverage}
			\end{figure}

			Fig. \ref{fig: coverage} compares perceived coverage given $\mathrm{QoE}_\mathrm{th}$ against traditional coverage, where the transmission rate is fixed.  The perceived coverage  is upper bounded by $r_{\max}=1$, and  lower bounded by the Rate-SNR function:
			
			{\begin{equation}
					\underline{r}(\mathrm{snr}) \triangleq \min r \,\, \mathrm{s.t.} \,\, \mathrm{QoE}(\mathrm{snr}, r) \le \mathrm{QoE}_{\mathrm{th}},
			\end{equation}}with two typical limit points: $\mathcal C_{\mathrm{limit}}^{\mathrm{snr}}=({\mathrm{snr}}^{\vdash},\underline{r}_{\mathrm{max}})$ and $\mathcal C_{\mathrm{limit}}^{\mathrm{rate}}=({\mathrm{snr}}^{\dashv},\underline{r}_{\mathrm{min}})$. ${\mathrm{snr}}^{\vdash}$ and ${\mathrm{snr}}^{\dashv}$  are the minimum required SNRs to guarantee the QoE performance given $\underline{r}_{\max}$ and $\underline{r}_{\min}$, respectively. The perceived coverage reveals joint dependencies on compression rates and channel conditions. In contrast, conventional coverage only relates to the channel conditions,  characterized by the limit point of $\mathcal C_{\mathrm{limit}}^{\mathrm{conv}}=({\mathrm{snr}}_{\mathrm{conv}}^{\vdash},\underline{r}_{\mathrm{conv}})$.

		 Perceived coverage can be significantly extended beyond conventional coverage due to the error-resilience of LightCom in deploying GenAI-augmented receiver. This coverage gain can be characterized in both power and bandwidth aspects:
			{\begin{equation}
					G(\mathrm{snr}, r) = G_\mathrm{power}(r) + G_\mathrm{bw}(\mathrm{snr}),
			\end{equation}}where $
				G_\mathrm{power}(r) = 10\log_{10}( {\mathrm{snr}_\mathrm{conv}}/{\mathrm{snr}})$ and  
				$G_\mathrm{bw}(\mathrm{snr}) = 10\log_{10}({r}/{r_\mathrm{conv}})$. Here, $\mathrm{snr}$ and $\mathrm{snr}_\mathrm{conv}$ denote the required SNRs to achieve the QoE requirement at a given rate $r$, while $r$ and $r_\mathrm{conv}$ represent the required compression rates at a given $\mathrm{snr}$, using the proposed and conventional frameworks, respectively.

			\section{Importance-Aware Power Allocation}
			It is challenging to derive an explicit expression of the QoE function in \eqref{eq:QoE}, making it difficult to achieve optimal QoE performance through resource allocation. However, as analyzed, the  resilience of the proposed LightCom can be approximately characterized through the tolerate IMSE, motivating us instead to use it as a objective function for sub-optimal resource allocation solutions. 

			Given the compression rate, we consider both uncoded and coded transmissions.  The problem in optimizing the QoE can be approximately formulated to minimize the IMSE as:
			
			\begin{subequations}\label{eq:Prob_bit_level}
				\begin{align}
					 \min_{p_k} \quad &\sum_{k=1}^{K}\gamma_k\mathrm{ber}_{k}\\
					\mathrm{s.t.} \quad & \sum_{k=1}^{K} p_k \le P, \label{eq:cons_power}
				\end{align}
			\end{subequations}where $p_k$ is the power allocated to each symbol of the $k$-th sub-stream. $P$ is the total power budget, and  $\mathrm{ber}_k \in \{\mathrm{ber}_k^{\mathrm u},  \mathrm{ber}_k^{\mathrm c}\}$ is the BER of the $k$-th sub-stream, with $\mathrm{ber}_k^{\mathrm u}$ and  $\mathrm{ber}_k^{\mathrm c}$ denoting the BERs under uncoded and coded transmission, respectively.
			 
			Problem \eqref{eq:Prob_bit_level} is convex w.r.t.  the allocated power $p_k$ due to the convexity of the BER functions in both channel-uncoded (i.e., \eqref{eq:BER_SNR_Uncoded}) and channel-coded cases (i.e., \eqref{eq:BER_SNR_Coded}).  An optimal solution exists and can be obtained via the Lagrange multiplier technique \cite{boyd2004convex}.  The  corresponding Lagrange function is given by: 	
			\begin{equation}
				\mathcal L(p_k, \lambda) \triangleq \sum_{k=1}^{K}\gamma_k\mathrm{ber}_{k} + \lambda\Big(\sum_{k=1}^{K} p_k -P\Big),
			\end{equation}where $\lambda$ is the Lagrange multiplier. The optimal solution satisfy:
			\begin{equation}\label{eq:lagrange}
				\frac{\partial \mathcal L(p_k, \lambda)}{\partial p_k}=\gamma_k\frac{\partial \mathrm{ber}_{k}}{\partial p_k} + \lambda=0.
			\end{equation}
			
			\subsubsection{Channel-uncoded} The gradient $\frac{\partial \mathrm{ber}_{k}}{\partial p_k}$  is yielded as:  
			

		{ 	\begin{align}\label{eq:gradient_u}
				\frac{\partial \mathcal \mathrm{ber}_{k}}{\partial p_k} = -\frac{1}{2\sqrt{\mathrm{snr}_k} }	\frac{\alpha_{\mathrm u}\beta_{\mathrm u}\vert h_k \vert^2}{\sqrt{2\pi}\sigma^2}\exp\Big(-\frac{\beta_{\mathrm u}^2 \mathrm{snr}_k}{2}\Big).
			\end{align}Substituting \eqref{eq:gradient_u} back into \eqref{eq:lagrange}, we obtain: 
				\begin{align}\label{eq:gradient}
				   	\frac{\alpha_{\mathrm u}\beta_{\mathrm u}\vert h_k \vert^2}{\sqrt{2\pi}\sigma^2}\exp\Big(-\frac{\beta_{\mathrm u}^2 \mathrm{snr}_k}{2}\Big)=\frac{2\lambda}{\gamma_k}\sqrt{\mathrm{snr}_k},
			\end{align}which is a transcendental equation. Due to the convexity of problem  \eqref{eq:Prob_bit_level}, the optimal solution satisfies the equality of constraint \eqref{eq:cons_power}.  However, there  is  not closed-form solution for $p_k$ given the the Lagrange multiplier $\lambda$,  necessitating numerical methods. One approach would be to search for the optimal $\lambda^*$ that satisfies the equality of \eqref{eq:cons_power}, but this would incur high computational complexity due to both numerical equation solving and Lagrange multiplier searching. 
			
			To address this challenge, we propose a low-complexity sub-optimal method by relaxing  the equation \eqref{eq:gradient} into:
			\begin{align}\label{eq:gradient_u_appr}
				\frac{\alpha_{\mathrm u}\beta_{\mathrm u}\vert h_k \vert^2}{\sqrt{2\pi}\sigma^2}\exp\Big(-\frac{\beta_{\mathrm u}^2 \mathrm{snr}_k}{2}\Big)=\frac{2\lambda}{\gamma_k}\sqrt{\widetilde{\mathrm{snr}}_k},
			\end{align}where $\widetilde{\mathrm{snr}}_k$ is approximated as $\widetilde{\mathrm{snr}}_k=\frac{\tilde{p}_k\vert h_k \vert^2}{\sigma^2}$, with  $\tilde{p}_k \triangleq \gamma_k \frac{P}{\sum_k \gamma_k}$. Since the allocated power cannot be negative, the yielded solution is then given by:

			\begin{align}\label{eq:allop_bit_level_appr}
				&p_k^* = \left( -\frac{2\sigma^2}{\beta_{\mathrm u}^2\vert h_k \vert^2}\ln \frac{2\lambda^*\sqrt{2\pi}\sigma^2\sqrt{\widetilde{\mathrm{snr}}_k}}{ \alpha_{\mathrm u}\beta_{\mathrm u} \gamma_k\vert h_k \vert^2} \right)^+\nonumber \\
				& = \underbrace{\frac{2\sigma^2}{\beta_{\mathrm u}^2\vert h_k \vert^2}}_{W_k^{\mathrm u}}\Bigg(\underbrace{\ln \frac{\alpha_{\mathrm u}\beta_{\mathrm u}}{2\sqrt{2\pi}\lambda^*} }_{H^{\mathrm u*}_\mathrm{level}} - \underbrace{\ln\frac{\sigma^2\sqrt{\widetilde{\mathrm{snr}}_k}}{\gamma_k\vert h_k \vert^2}}_{H_k^{\mathrm u}} \Bigg)^+,
			\end{align}where $(\cdot)^+$ denotes the  $\max(0,\cdot)$ operation, and $\lambda^*$ is the optimal Lagrange multiplier. This forms the WF solution, where $W_{k}^{\mathrm u}$ and $H_{k}^{\mathrm u}$ can be interpreted as the base widths and heights, respectively. $H_\mathrm{level}^{^{\mathrm u*}}$ represents the optimal water level corresponding to $\lambda^*$, which satisfies the equality of power constraint (\ref{eq:cons_power}):

			 \begin{equation}
			 	\sum_{k=1}^{K} p_k^* = \sum_{k=1}^{K} W_k^{\mathrm u} \left({H^{{\mathrm u*}}_\mathrm{level}} - {H_k^{\mathrm u}} \right)^+ = P.
			 \end{equation} 
			 
		$H_\mathrm{level}^{\mathrm u*}$ can be solved using the bisection search technique.  Denote the sum of the allocated power as a function of ${H}_\mathrm{level}^{\mathrm u}$, i.e., $
			f({H}_\mathrm{level}^{\mathrm u}) = \sum_{k=1}^{K} W_k^{\mathrm u} \left({H^{{\mathrm u}}_\mathrm{level}} - {H_k^{\mathrm u}} \right)^+$.
	 It is monotonically increasing with ${H}_\mathrm{level}^{\mathrm u}$. To implement the bisection search, we must first establish appropriate lower and upper bounds for the water level. For the lower bound $\underline{H}_\mathrm{level}^{\mathrm u}$,  we need to ensure $f(\underline{H}_\mathrm{level}^{\mathrm u})<P$. Given the function's properties, setting $\underline{H}_\mathrm{level}^{\mathrm u} \triangleq \min_k(H_k^{\mathrm u})$  guarantees that $f(\underline{H}_\mathrm{level}^{\mathrm u})=0<P$. For the upper bound $\overline{H}_\mathrm{level}^{\mathrm u}$,  we need $\overline{H}_\mathrm{level}^{\mathrm u}>\max_k({H_k^{\mathrm u}})$ and must ensure $f( \overline{H}_\mathrm{level}^{\mathrm u})\ge P$. This can be achieved by satisfying:
		\begin{align}
			&f( \overline{H}_\mathrm{level}^{\mathrm u})\ge  \min_k (W_k^{\mathrm u}) \sum_{k=1}^{K}  ({\overline{H}^{{\mathrm u}}_\mathrm{level}} - {H_k^{\mathrm u}}),\nonumber\\
			& \ge  K \min_k (W_k^{\mathrm u}) ({\overline{H}^{{\mathrm u}}_\mathrm{level}} - \max_k({H_k^{\mathrm u}}))\ge P.
		\end{align}Therefore, the upper bound can be set to $\overline{H}_\mathrm{level}^{\mathrm u}\triangleq P/(K\min_k (W_k^{\mathrm u}) )+\max_k({H_k^{\mathrm u}})$. With these bounds established, the optimal power allocation can be determined via bisection search, as summarized in \textbf{Algorithm}  \ref{alg:alg1}.	
	}

		\begin{algorithm}[t] 
			\renewcommand{\algorithmicrequire}{\textbf{Input:}}
			\renewcommand{\algorithmicensure}{\textbf{Output:}}
			\caption{Importance-Aware Waterfilling for Channel-Uncoded Case.} 
			\label{alg:alg1}
			\begin{algorithmic}[1]		
				
				\STATE  Initialize the lower and upper bounds: $\underline {H}_{\mathrm{level}}^{\mathrm u}=\min_k(H_k^{\mathrm u})$, $\overline {H}_{\mathrm{level}}^{\mathrm u}=P/(K\min_k (W_k^{\mathrm u}) )+\max_k({H_k^{\mathrm u}})$. 
				
				\STATE \textbf{While} $\overline {H}_{\mathrm{level}}^{\mathrm u} - \underline {H}_{\mathrm{level}}^{\mathrm u}\ge \epsilon$
				\STATE  \quad Set new $ {H}_{\mathrm{level}}^{\mathrm u} = (\overline {H}_{\mathrm{level}}^{\mathrm u} + \underline {H}_{\mathrm{level}}^{\mathrm u})/2$
				\STATE  \quad Compute $p_k$ based on (\ref{eq:allop_bit_level_appr})	
				\STATE  \quad \textbf{If} $\sum_{k=1}^{K}p_k \le P$
				\STATE  \quad \quad Update $\underline {H}_{\mathrm{level}}^{\mathrm u}\leftarrow {H}_{\mathrm{level}}^{\mathrm u}$
				\STATE  \quad \textbf{Else}
				\STATE  \quad \quad  Update $\overline {H}_{\mathrm{level}}^{\mathrm u}\leftarrow {H}_{\mathrm{level}}^{\mathrm u}$
				
				\STATE  \quad \textbf{End}	
				\STATE  \textbf{End}
				\STATE   Compute $p_k^*$ based on (\ref{eq:allop_bit_level_appr})	
			\end{algorithmic}  
			
		\end{algorithm}

		\subsubsection{Channel-coded}The power allocation strategy also can be optimized   using a similar approach. According to our previous work \cite{xu2025dataimportanceJ} and due to the equal length of sub-streams, the optimal power allocation is given by:
			\begin{equation}\label{eq:allop_coded}
				p_k^* = \underbrace{-\frac{\sigma^2}{\beta_{\mathrm c} \vert h_k \vert^2}}_{W_k^{\mathrm c}}\Big(\underbrace{\ln \frac{-\alpha_{\mathrm c}\beta_{\mathrm c}}{\lambda^*} }_{H^{\mathrm c*}_\mathrm{level}} - \underbrace{\ln\frac{\sigma^2}{\gamma_k\vert h_k \vert^2}}_{H_k^{\mathrm c}} \Big)^+.
			\end{equation} $H_\mathrm{level}^{\mathrm c*}$ represents the optimal water level corresponding to the optimal $\lambda^*$, satisfying the equality of constraint (\ref{eq:cons_power}):
			\begin{equation}
				\sum_{k=1}^{K} {W_k^{\mathrm c}}\left({H^{\mathrm c*}_\mathrm{level}} - {H_k^{\mathrm c}} \right)^+ = P.
			\end{equation} $H_\mathrm{level}^*$ can be solved based on the bisection search technique. Similarly, the lower bound of $H_\mathrm{level}^*$ can be set to
			$\underline{H}_\mathrm{level}^{\mathrm c} = \min_k(H_k^{\mathrm c})$, which ensures that  $\sum_{k=1}^{K} {W_k^{\mathrm c}}\left({H^{\mathrm c*}_\mathrm{level}} - {H_k^{\mathrm c}} \right)^+=0<P$. The upper bound  of $H_\mathrm{level}^*$ can be set to  $\overline{H}_\mathrm{level}^{\mathrm u}=P/(\min_k (W_k^{\mathrm u}) K)+\max_k({H_k^{\mathrm u}})$ to guarantee $\sum_{k=1}^{K} {W_k^{\mathrm c}}\left({H^{\mathrm c*}_\mathrm{level}} - {H_k^{\mathrm c}} \right)^+>P$.  The procedure for solving the optimal power allocation $p_k^*$  for the channel-coded case follows a similar pattern to that of the channel-uncoded case, with appropriate adjustments to the water level and base height calculations.

			\section{Simulation Results}
		 	\begin{figure*}[tp]
		 	\centering
		 	\includegraphics[width=1\textwidth]{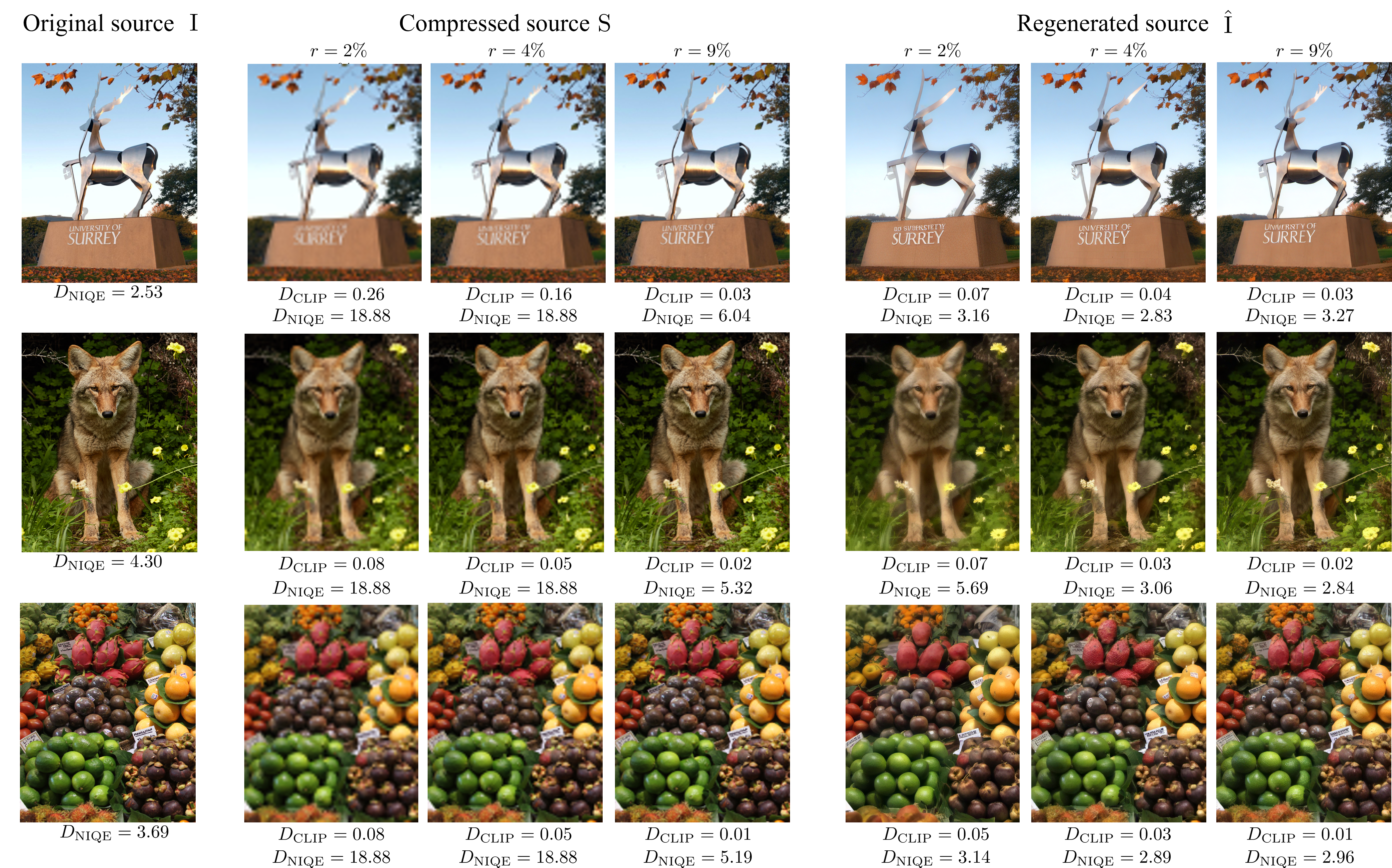}
		 	\caption{Visual results of the original source, compressed representation obtained by \re{LPF} encoding, and the reconstructed content by the GenAI-augmented decoding under the proposed LightCom framework.}
		 	\label{fig:sourceCoding}
		 \end{figure*}
		 
		 \begin{figure}[tp]
		 	\centering
		 	\subfigure[]{\includegraphics[width=0.23\textwidth]{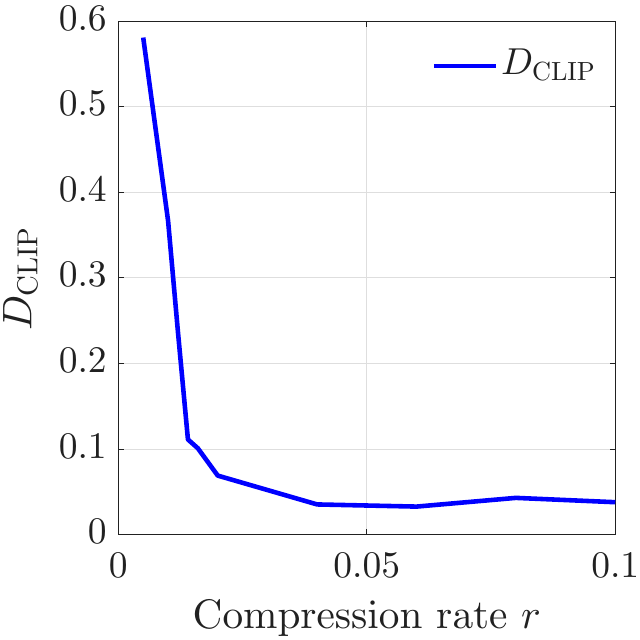}}
		 	\hspace{5pt}
		 	\subfigure[]{\includegraphics[width=0.23\textwidth]{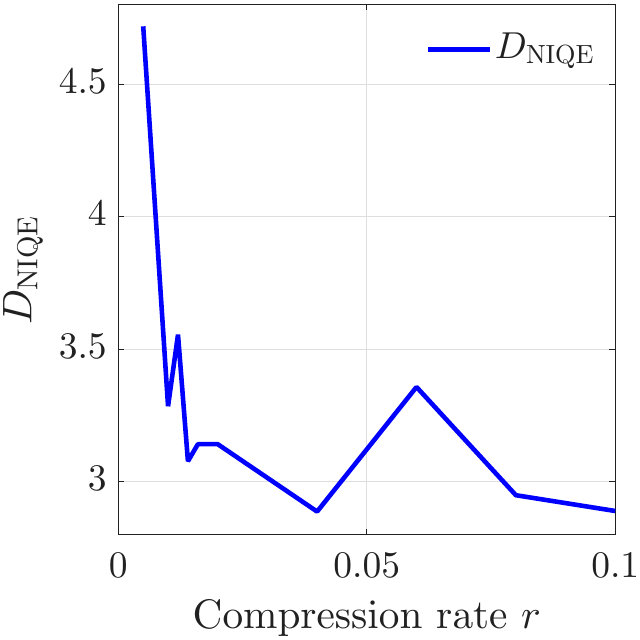}}
		 	\caption{The QoE performance versus compression rates under error-free transmission.}
		 	\label{fig:sufficiency}
		 \end{figure}
		 \begin{figure}[tp]
		 	\centering
		 	\subfigure[]{\includegraphics[width=0.23\textwidth]{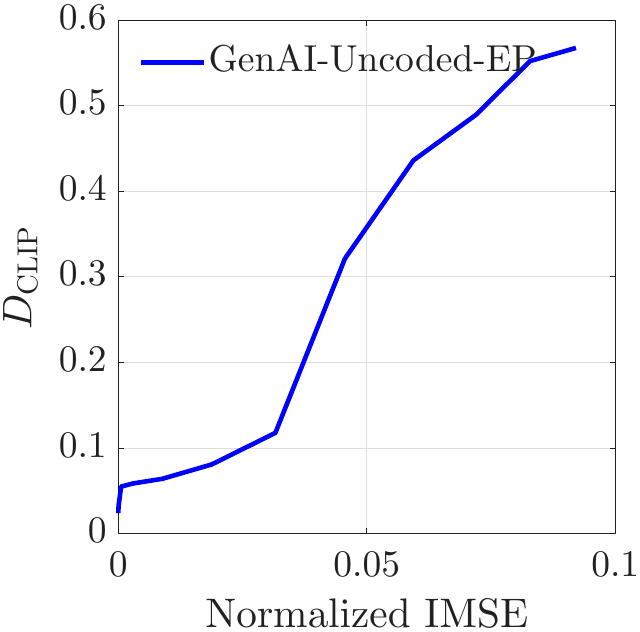}}
		 	\hspace{5pt}
		 	\subfigure[]{\includegraphics[width=0.23\textwidth]{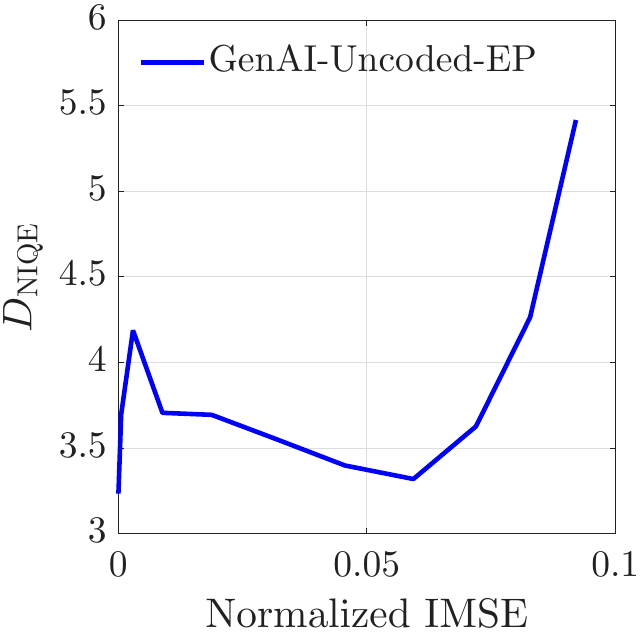}}
		 	
		 	\caption{The QoE performance versus the normalized MSE  at a compression rate of $9\%$.}
		 	\label{fig:explicity}
		 \end{figure}

				This section presents simulation results  validating the effectiveness of the proposed LightCom framework for QoE-oriented communications. Three RGB images of size $H=640$, $W=512$, and $K=8$-bit depth are selected to represent diverse visual characteristics: (i) clear object with background, (ii) fine-grained textures, and (iii) rich object without background. The source encoder applies a low-pass mean filter with a uniform kernel, where the value corresponds to the compression rate \( r \). At the receiver, the trained generative IR model, i.e., SUPIR model,is employed for  reconstruction, leveraging its strong generative capabilities and support for text-guided enhancement \cite{yu2024scaling}.  
				
				For importance-aware transmission implementation, the compressed source is partitioned into $K=8$ equal-length bit sequences according to bit-level importance.  Each bit sequence is  protected using WCC, with both no coding and $(7,4)$ Hamming coding considered, and subsequently modulated using QPSK. Power is allocated to each sub-stream using the proposed importance-aware WF methods. Since orthogonal sub-channels with varying gains can be equivalently modeled as AWGN channels with different noise variances, we consider AWGN channels with unit variance in this section. The QoE requirement is set to $\mathrm{QoE}_\mathrm{th}\triangleq [D_\mathrm{NIQE}^\mathrm{th}, D_\mathrm{CLIP}^\mathrm{th}]=[5, 0.1]$.  QoE performance results are obtained by  averaging over $100$ independent realizations, with a single representative image used.  
				For baseline comparison, we implement current communication systems using JPEG compression and capacity-approaching LDPC codes in channel-coded scenarios. 	Since conventional WF for power allocation reduces to equal power (EP) strategies under AWGN channels, we employ EP allocation for the baselines. 
				
				We compare the following approaches:
				\begin{itemize}
					\item  GenAI-Uncoded-WF/GenAI-Uncoded-EP: Our LightCom framework with uncoded transmission using importance-aware WF  or  EP allocation.
	 
					\item GenAI-Hamming-WF/GenAI-Hamming-EP: Our LightCom framework with Hamming code   using importance-aware WF or EP allocation.
			 
					\item GenAI-Conv.-WF/GenAI-Conv.-EP: Our LightCom framework with  convolutional code   using importance-aware WF or  EP allocation.
			 
					\item  JPEG-Uncoded-EP/JPEG-LDPC-EP: Traditional communication baselines without channel coding or with LDPC.
					
				\end{itemize}
			It is worth noting that deep JSCC baselines are not included for comparison. The reasons are twofold: Prior study have shown that separate source and channel coding schemes can achieve superior performance by leveraging the strength of the large language models \cite{ren2025separate}. Moreover, deep JSCC methods often exhibit limited generalization and high sensitivity to varying channel conditions, making them less robust under the uncertain and low-SNR settings similar to traditional baselines.
			
				  \begin{figure*}[tp]
				\centering
				\includegraphics[width=1\textwidth]{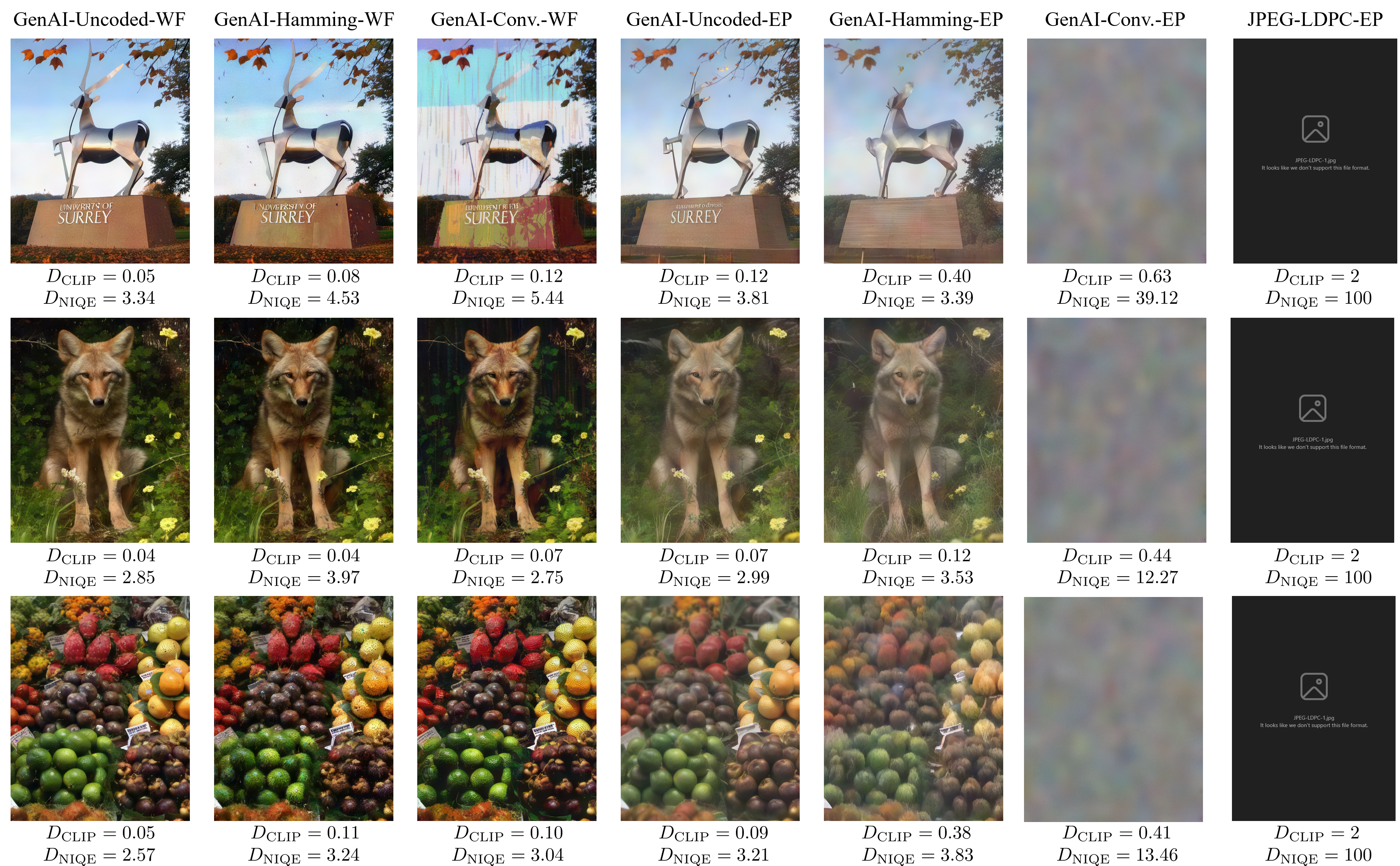}
				\caption{Visual results of the reconstructed sources across the proposed LightCom framework and traditional communication systems, where $r=9\%$ and $E_b/N_o=-2$~dB.}
				\label{fig:channelCoding}
			\end{figure*}

			\subsection{Effectiveness of the LightCom Framework} 
				
				Fig. \ref{fig:sourceCoding}  visualizes the original source, compressed representation obtained by \re{LPF} encoding, and the reconstructed content by the GenAI-augmented decoding, where contextual cues (e.g.,``University of Surrey'' ) guide reconstruction of the first image source. It shows the visual quality degradation of compressed sources as compression rate decreases, resulting in increasingly blurring images. Thanks to the powerful generative capabilities of the deployed GenAI model, coherent details are semantically inferred from the distorted signals, producing high-fidelity reconstructions with high QoE performance. Despite substantial visual distortion at a compression rate of $2\%$, the GenAI-augmented receiver semantically reconstructs perceptually faithful content. These results validate the effectiveness of the LightCom framework using the lightweight \re{LPF} source encoder paired with the generative source decoder in reconstructing QoE-satisfactory content.
				
				Figs. \ref{fig:sufficiency} and \ref{fig:explicity} examine the sufficiency and error-resilience of the source representation under the LightCom framework. Fig. \ref{fig:sufficiency} confirms that the QoE generally improves with the compression rate, and the  QoE requirement is met when  $r>2\%$. In contrast, Fig. \ref{fig:explicity} illustrates that QoE degrades with increasing residual errors, measured by normalized MSE, at a fixed compression rate of 9\%. Specifically, the QoE requirement is met when the normalized MSE remains below $-16$ dB under uncoded transmission with EP allocation. These results confirm the analysis pf both sufficiency and error-resilience of the source representation, which are crucial to achieving target QoE requirement.

	  \subsection{QoE Performance Comparison}

	  We present comprehensive QoE comparisons with traditional communication methods. Fig. \ref{fig:channelCoding} depicts visualized results of the reconstructed images at a compression rate of $r=9\%$ and SNR of $E_b/N_o=-2$ dB. The results show that the proposed LightCom with WCCs yield superior visual quality even without channel coding than the convolutional codes, whereas traditional baseline with strong LDPC completely fails at reconstruction.  Furthermore, the importance-aware WF approaches consistently deliver better visually perceived quality than EP allocation strategies across channel-uncoded, Hamming-coded, and convolutional-coded scenarios.

	\begin{figure*}[tp]
		\centering
		\subfigure[]{\includegraphics[width=0.31\textwidth]{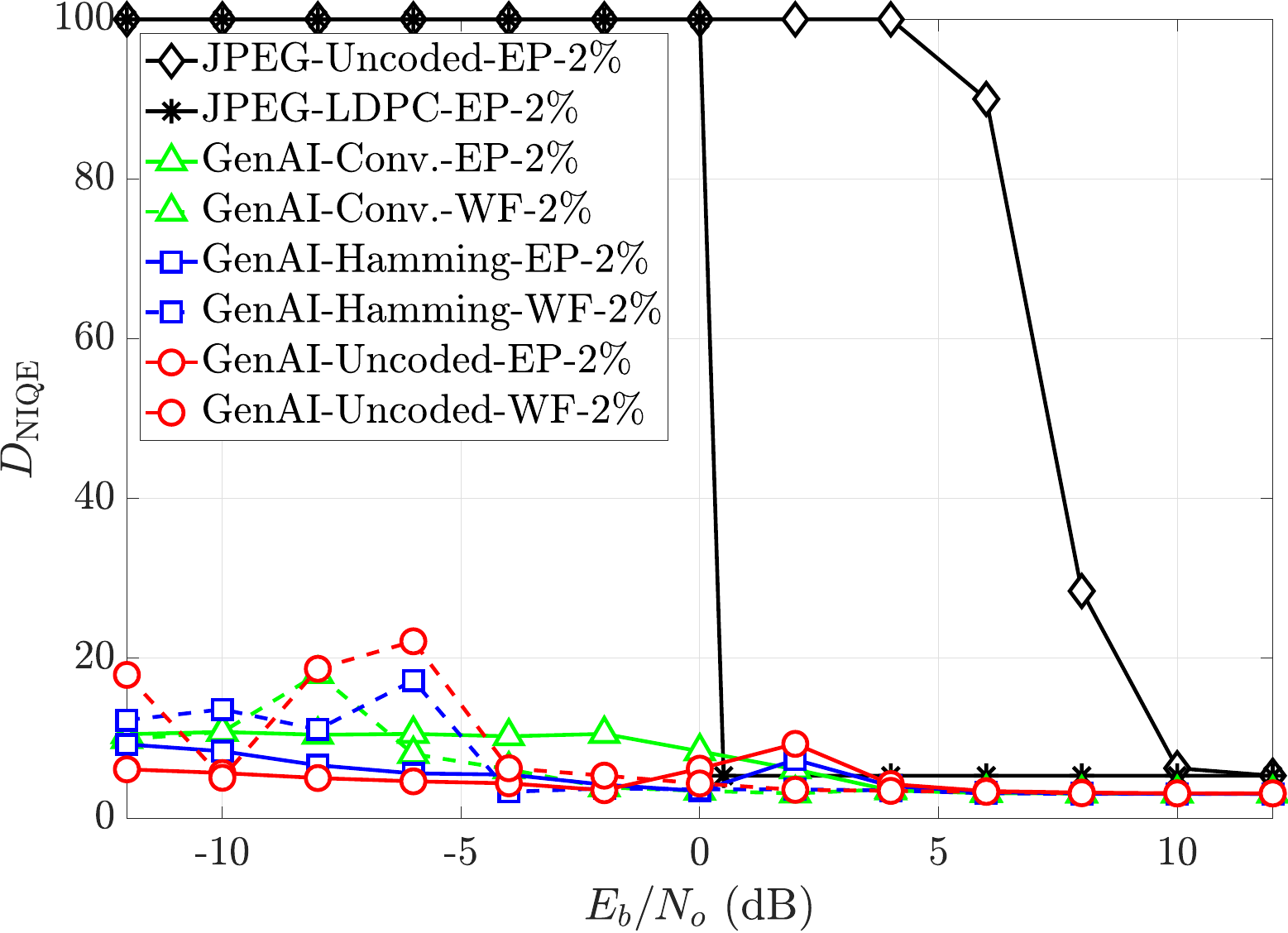}}
		\hspace{5pt}
		\subfigure[]{\includegraphics[width=0.31\textwidth]{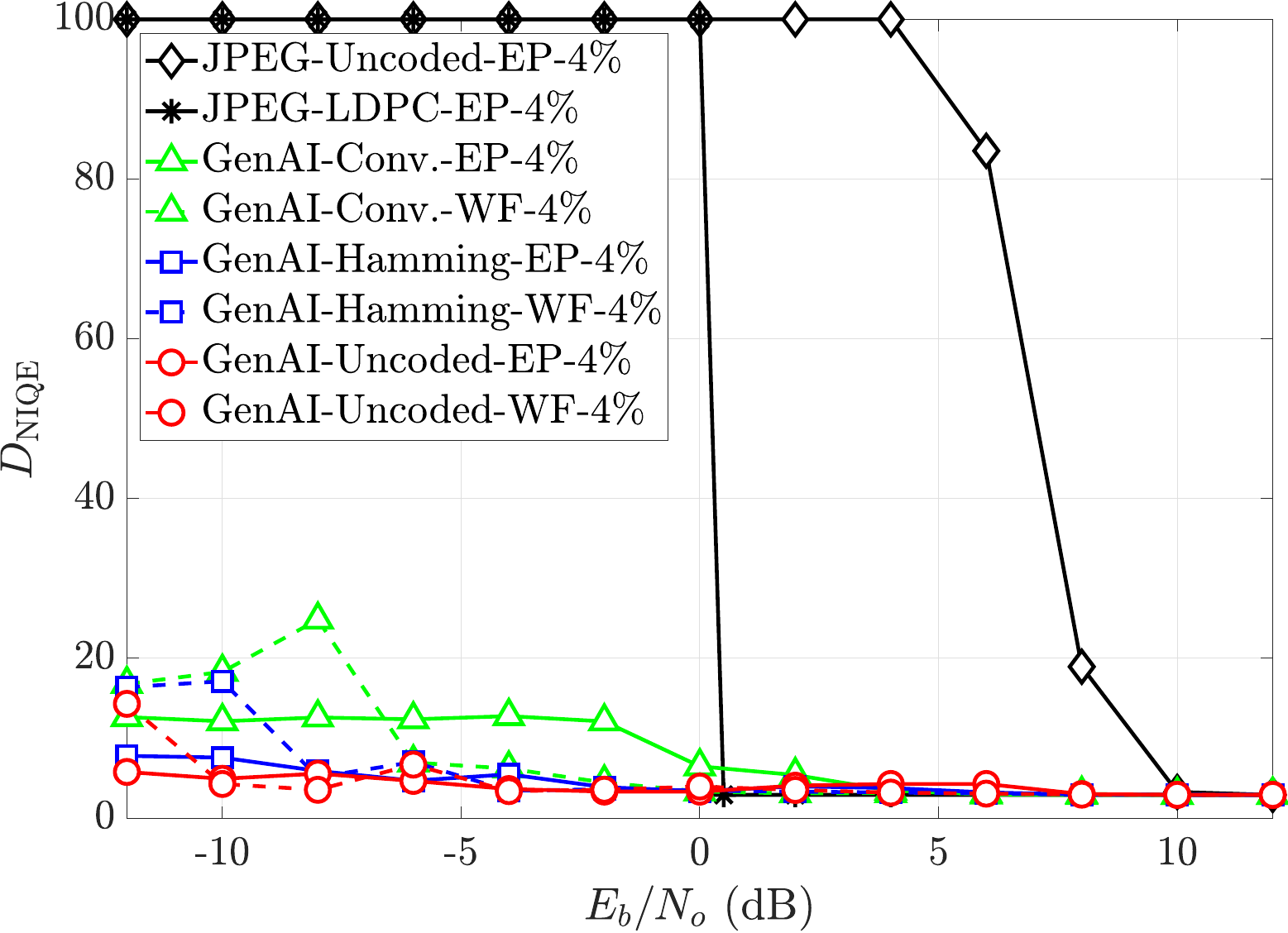}}
		\hspace{5pt}
		\subfigure[]{\includegraphics[width=0.31\textwidth]{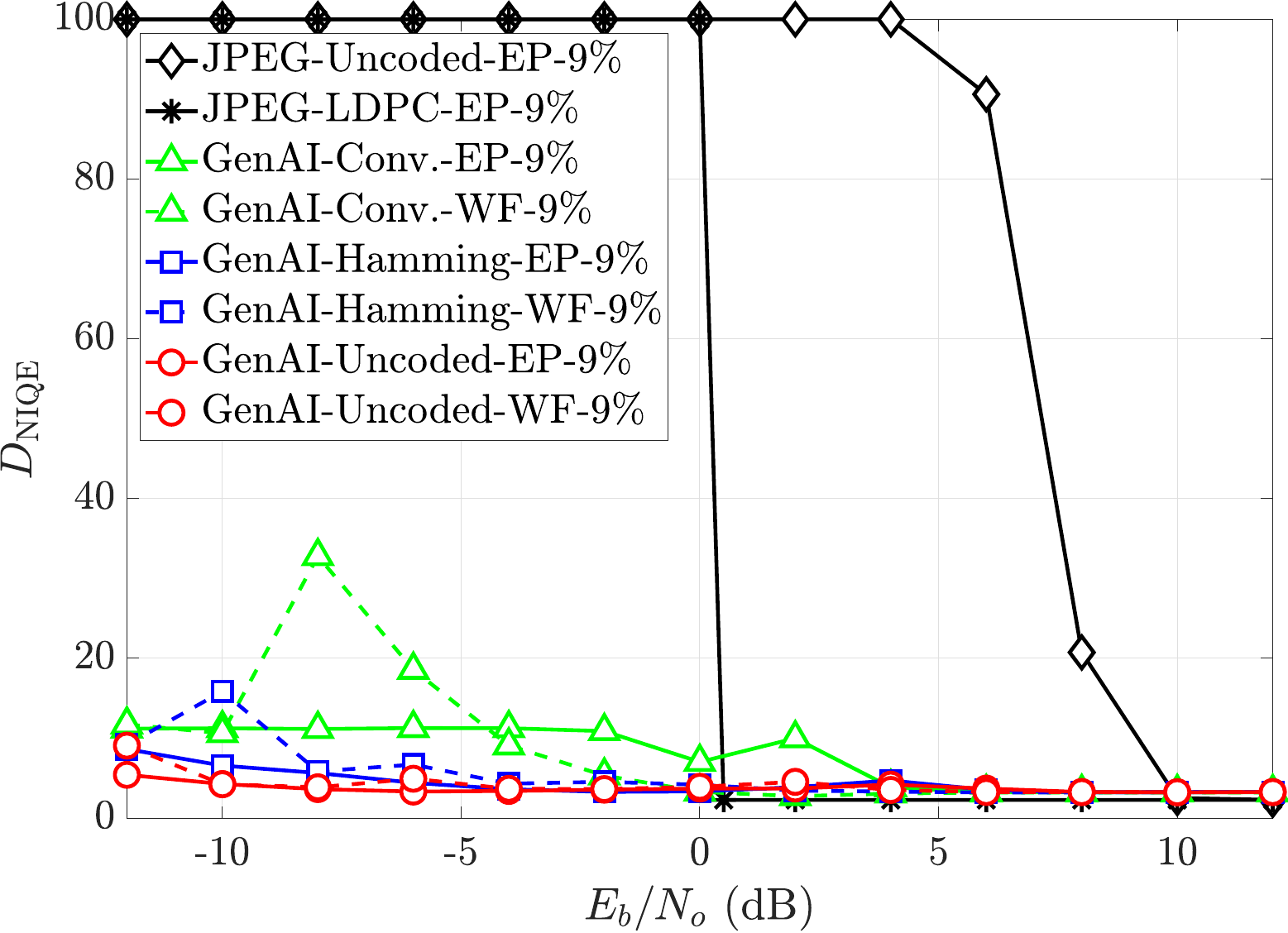}}

		\subfigure[]{\includegraphics[width=0.31\textwidth]{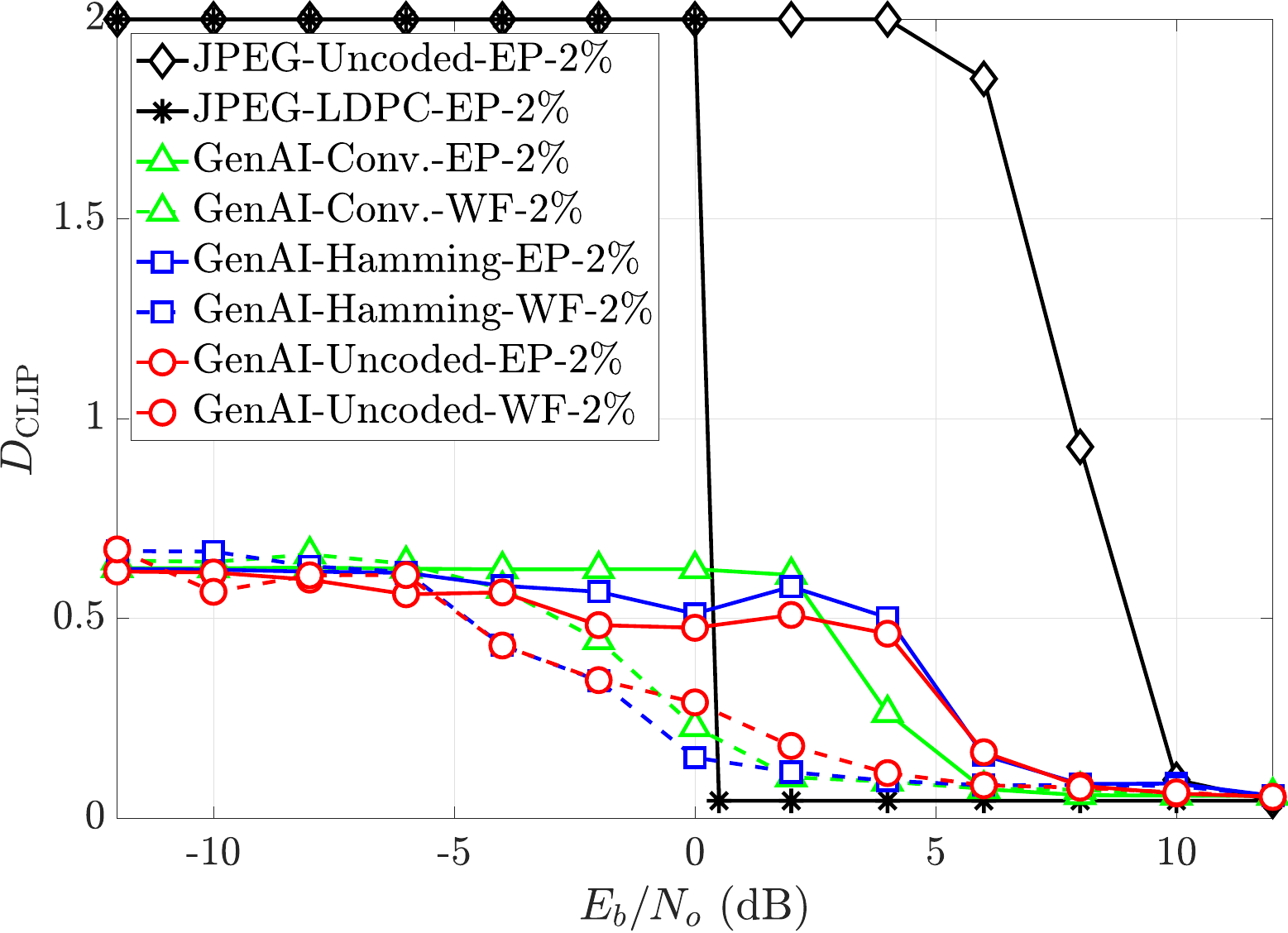}}
		\hspace{5pt}
		\subfigure[]{\includegraphics[width=0.31\textwidth]{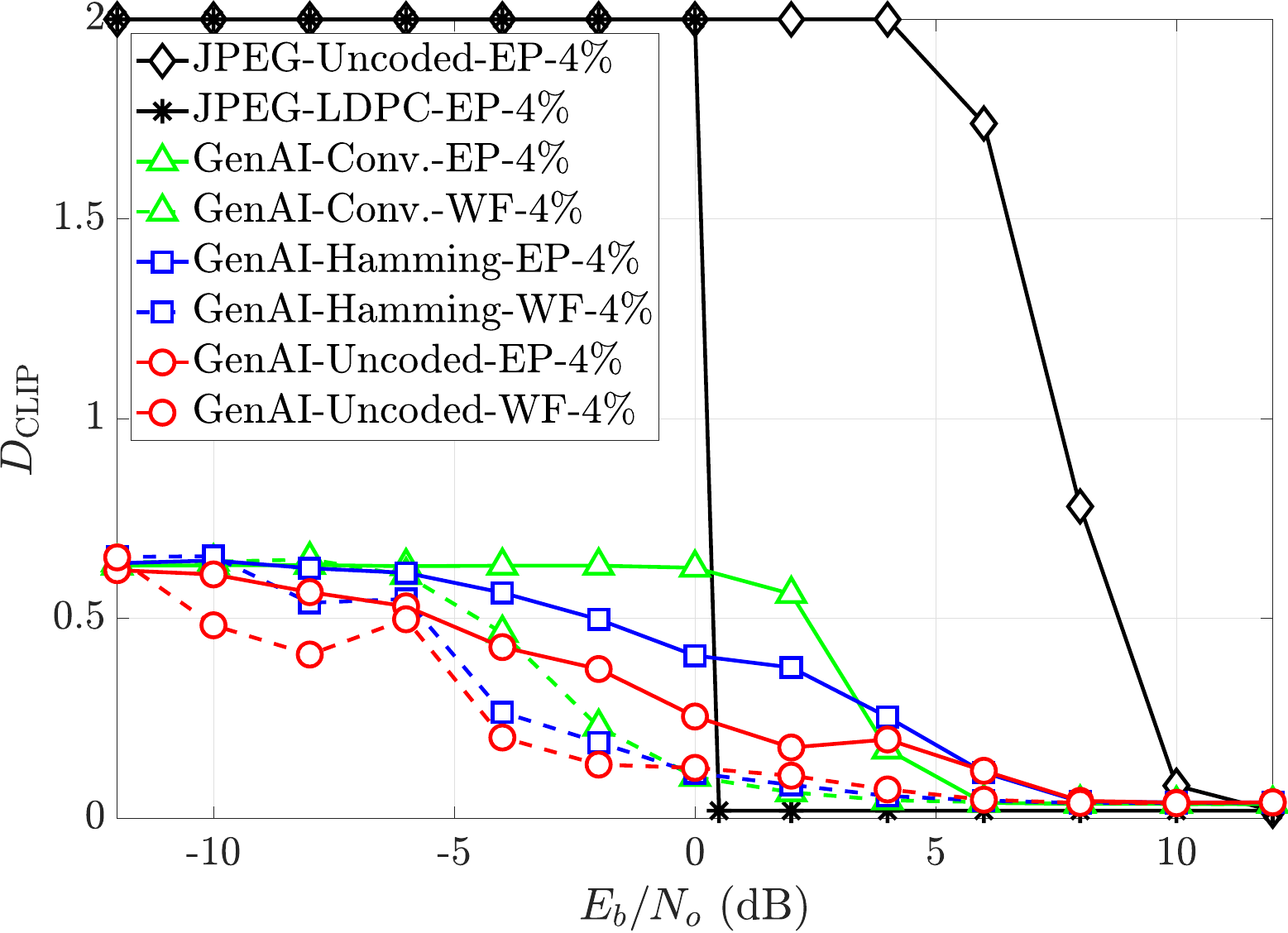}}
		\hspace{5pt}
		\subfigure[]{\includegraphics[width=0.31\textwidth]{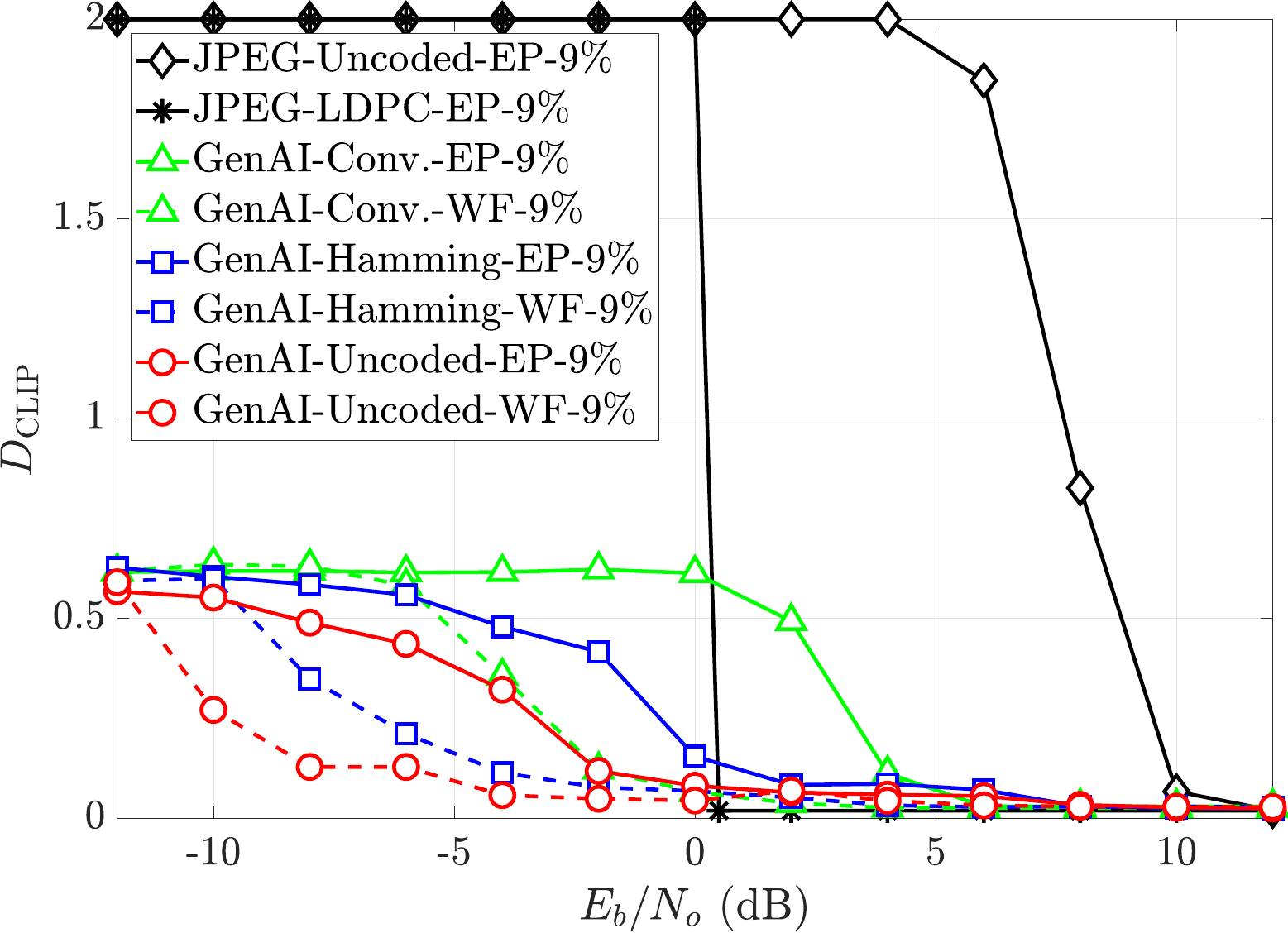}}
		
		\caption{QoE performance comparison under different compression rates. $D_\mathrm{NIQE}$: (a). $r=2\%$; (b).  $r=4\%$; (c). $r = 9\%$. $D_\mathrm{CLIP}$: (d). $r=2\%$; (e). $r=4\%$; (f). $r=9\%$. }
		\label{fig:QoE_Comparison}
	\end{figure*}
	
	\begin{figure*}[tp]
		\centering
		\subfigure[]{\includegraphics[width=0.31\textwidth]{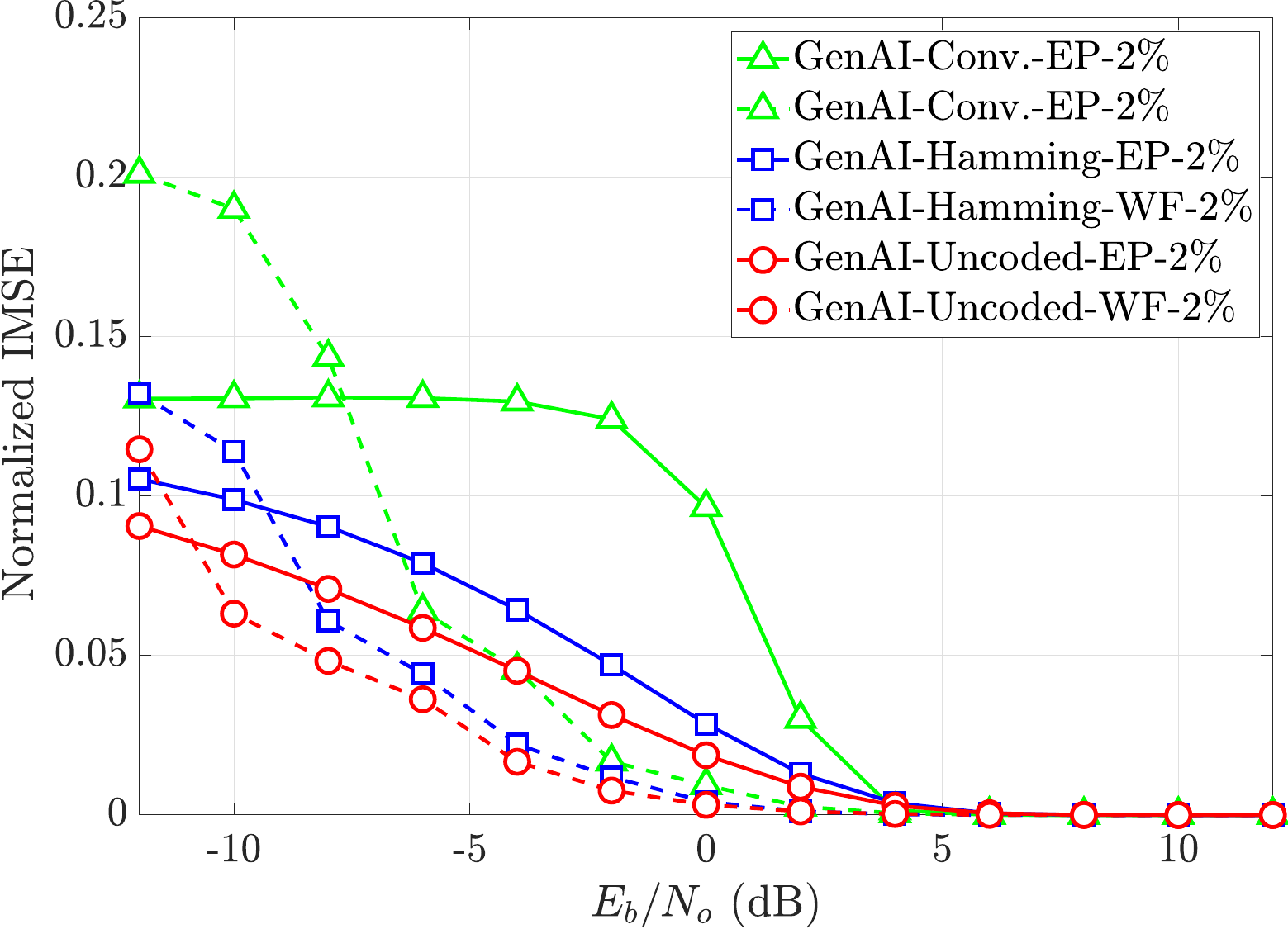}}
		\hspace{5pt}
		\subfigure[]{\includegraphics[width=0.31\textwidth]{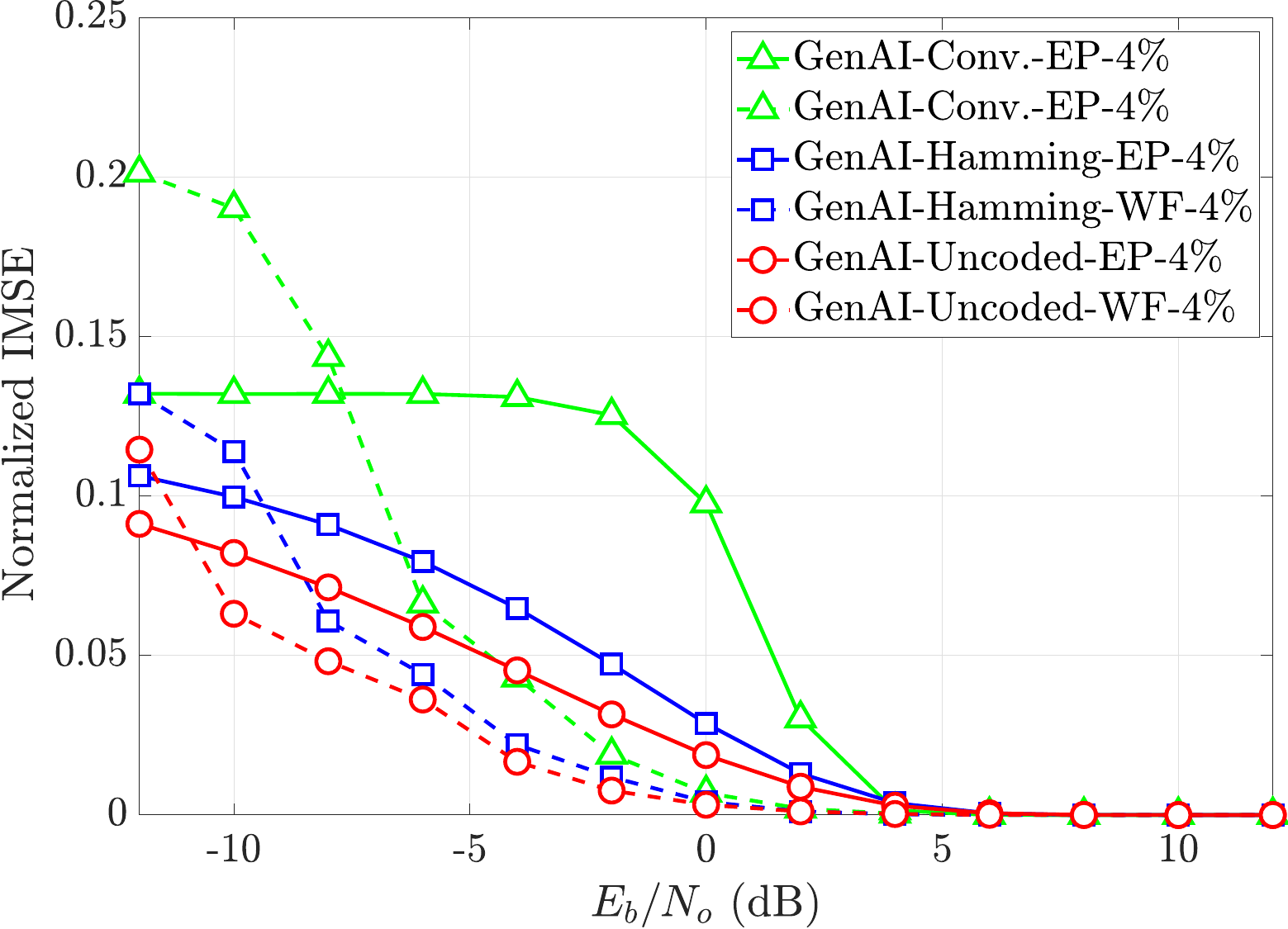}}
		\hspace{5pt}
		\subfigure[]{\includegraphics[width=0.31\textwidth]{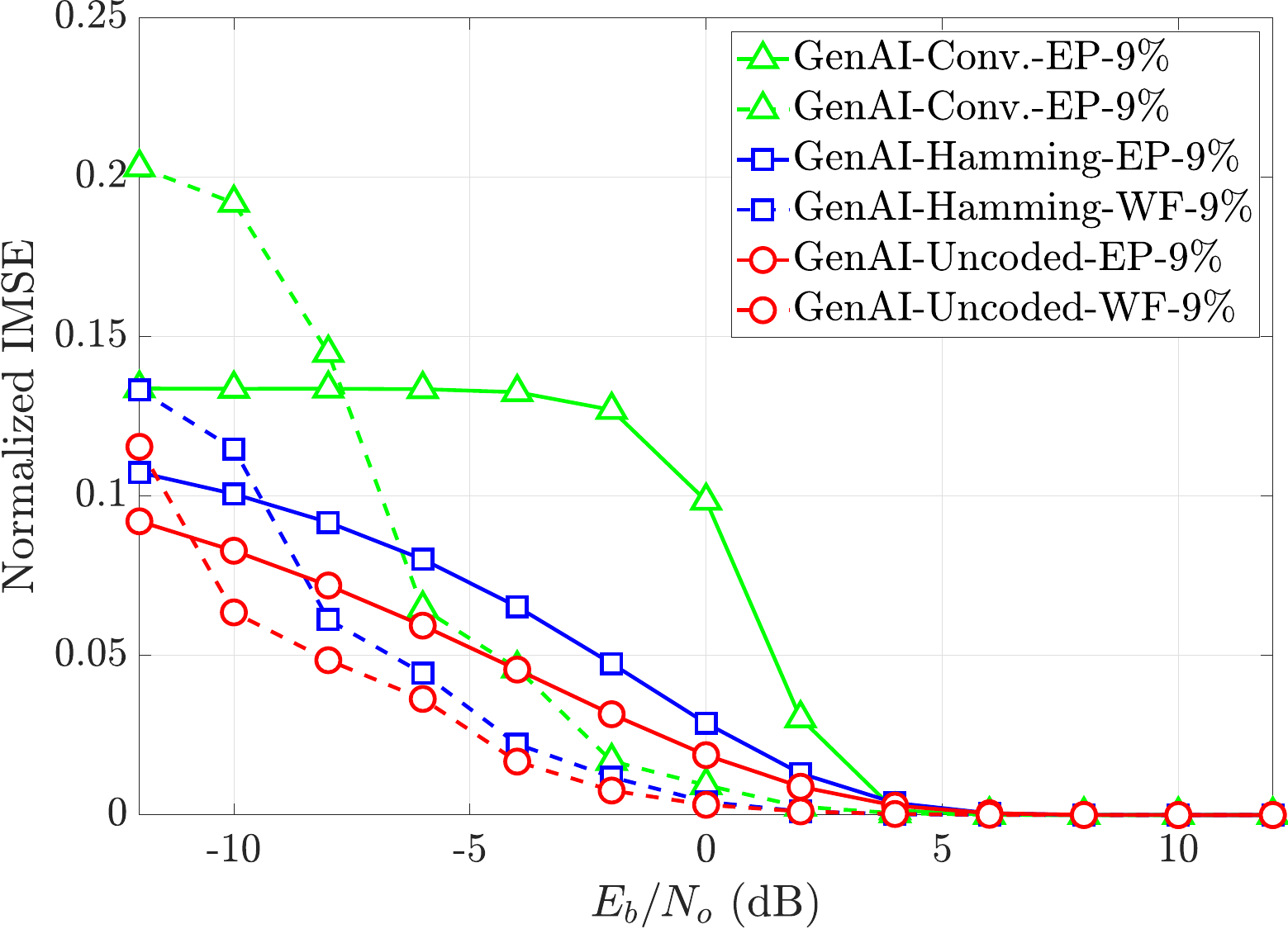}}
		\caption{The normalized IMSE comparison under different compression rates: (a). Compression rate of $2\% $; (b). Compression rate of $4\% $; (c) Compression rate $9\%$. 
		}
		\label{fig:MSE_SNR}
	\end{figure*} 
	   Fig. \ref{fig:QoE_Comparison}  illustrates QoE performance comparisons at different compression rates of $r=2\%$,  $r=4\%$, and  $r=9\%$. Traditional systems completely fail to meet QoE below  $E_b/N_o\le 4$ dB in the uncoded scenario and $E_b/N_o\le 0$ dB with LDPC coding. 	
	   Within the proposed LightCom framework, both $D_\mathrm{NIQE}$  and $D_\mathrm{CLIP}$  show generally decreasing trends as SNR increases  with some fluctuations. These fluctuations are attributed to distribution mismatches between the corrupted source representations and low-quality image datasets used for SUPIR model training.  At low SNRs, uncoded transmission achieves the best QoE performance, followed by Hamming coding and then convolutional coding. This ordering arises because the residual  errors become progressively more structured and correlated, making decoded representations less consistent with training datasets. However, at higher SNRs, convolutional coding ultimately surpasses other schemes, as transmission errors are significantly reduced with strong error protection. These findings underscore the importance of fine-tuning GenAI models on error-contaminated inputs to improve reconstruction robustness.

	  Additionally, within the proposed LightCom framework, increasing the compression rate leads to a reduction in the SNR required to meet the QoE requirement, in contrast to conventional systems where the required SNR remains largely invariant with fixed transmission rate. 	This implies that higher-rate compressed sources provide greater redundancy to enhance the GenAI model's semantic inference capability, improving robustness to channel impairments. These results suggest to design new lightweight channel coding schemes that incorporate source redundancy. Furthermore,  the proposed importance-aware WF approaches consistently outperforms the EP allocation strategies, particularly in terms of $D_\mathrm{CLIP}$ when SNR exceeds  certain threshold. At $r=9\%$, the required SNRs to guarantee the QoE requirement under the proposed GenAI-Uncoded-WF approach are reduced by $14$ dB and $4.5$ dB compared to the JPEG-Uncoded-EP and JPEG-LDPC-EP baselines, respectively. Notably, the  proposed WF strategies are designed to minimize MSE, with the normalized MSE illustrated in  Fig. \ref{fig:MSE_SNR}.  Results demonstrate that the proposed WF methods achieves lower MSE than  EP allocation strategies above certain SNR thresholds. Although not explicitly  optimized for QoE, MSE minimization yields sub-optimal yet effective solutions, demonstrating its practicality  a surrogate objective for resource allocation in the LightCom framework for QoE-oriented communications.

		\subsection{Perceived Coverage and Robustness}
		
\begin{figure}[tp]
				\centering
				\includegraphics[width=0.9\columnwidth]{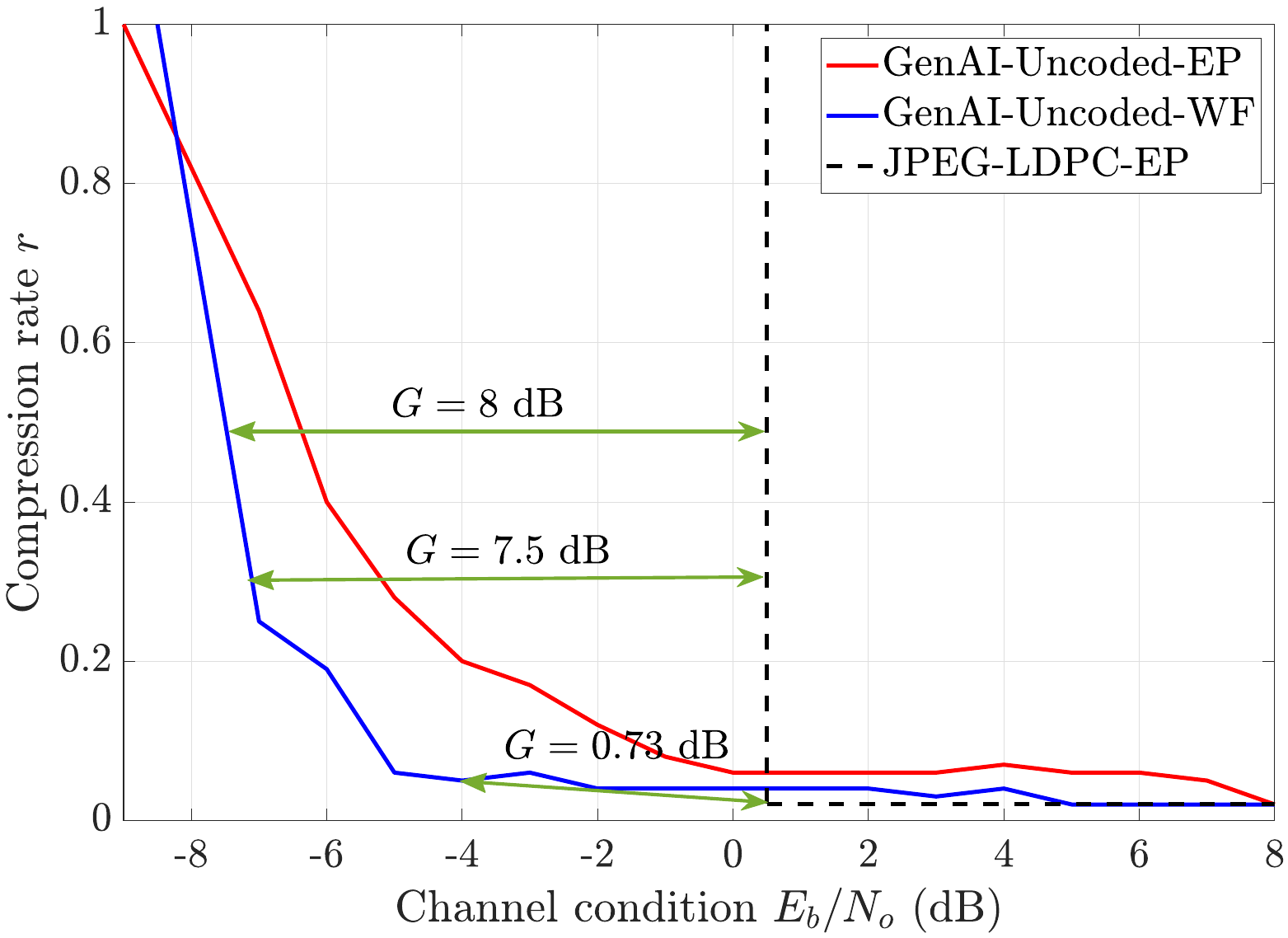}
				\caption{Perceived coverage of LightCom and conventional communication
					system with fixed transmission rates.}
				\label{fig:perceivedCoverage}
			\end{figure}

\begin{figure}[tp]
				\centering
				\includegraphics[width=1\columnwidth]{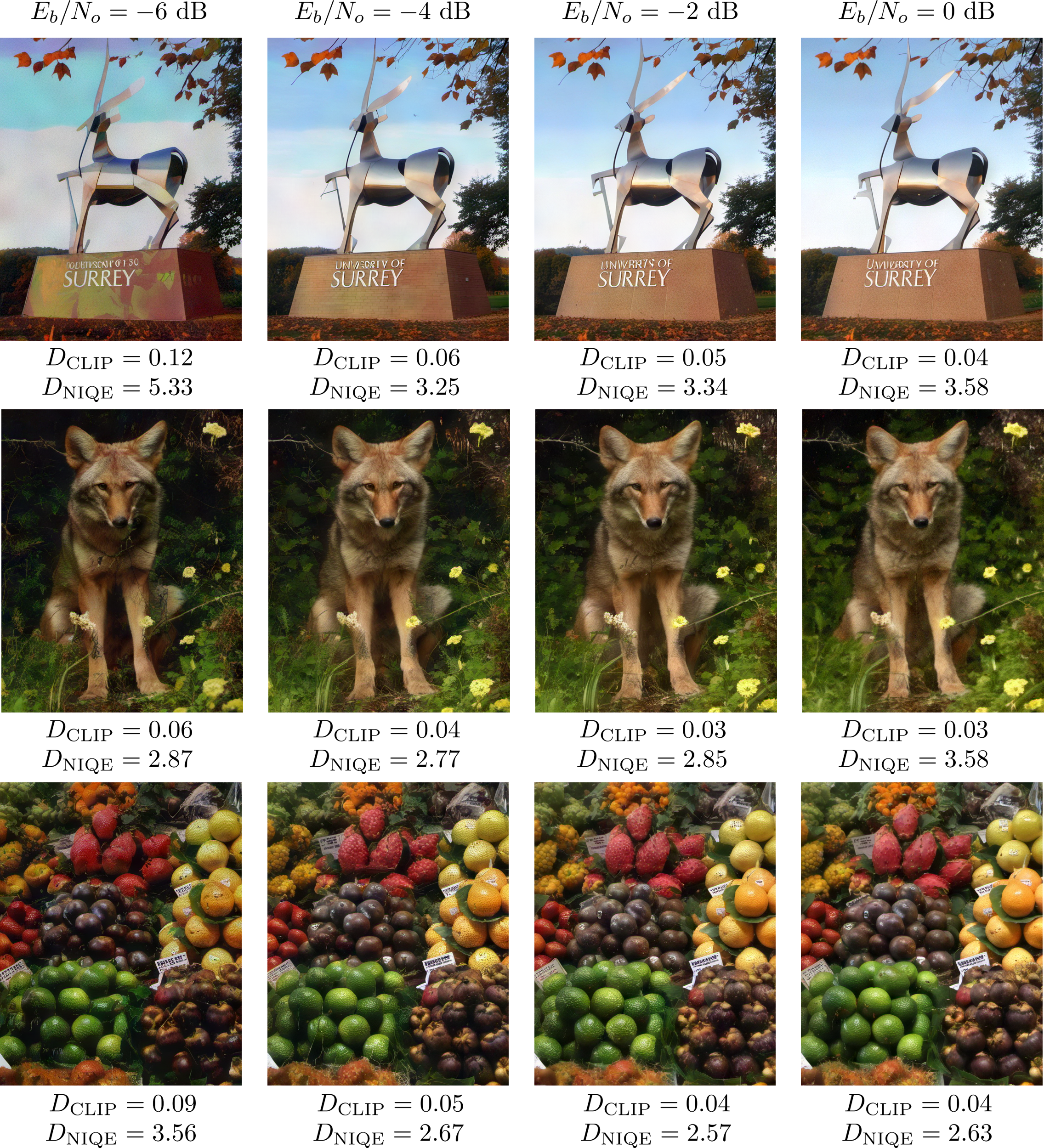}
				\caption{{Visualization of the reconstructed images at low SNRs using the proposed GenAI-Uncoded-WF approach at a compression rate of $r=9\%$.}}
				\label{fig:robustChannel}
			\end{figure}

		Since the lowest SNRs required to satisfy the QoE requirement are observed in the channel-uncoded LightCom and  LDPC-coded traditional systems,  Fig.~\ref{fig:perceivedCoverage} compares their perceived coverage under fixed transmission rates,  determined by the respective channel coding rates and modulation orders of each approach. For the conventional JPEG-LDPC-EP baseline, the QoE requirement is only satisfied when \( E_b/N_o \geq 0.5 \) dB and \( r \geq 2.1\% \), exhibiting a sharp threshold behavior with the limit point at \( \mathcal{C}_{\mathrm{limit}}^{\mathrm{conv}} = (0.5\,\mathrm{dB},\,2.1\%) \).  In contrast, the limit points under our LightCom framework are approximately \( \mathcal{C}_{\mathrm{limit}}^{\mathrm{snr}} = (-8.5\,\mathrm{dB},\,100\%) \) and \( \mathcal{C}_{\mathrm{limit}}^{\mathrm{rate}} = (5\,\mathrm{dB},\,2\%) \) using WF approach, and \( \mathcal{C}_{\mathrm{limit}}^{\mathrm{snr}} = (-9\,\mathrm{dB},\,100\%) \) and \( \mathcal{C}_{\mathrm{limit}}^{\mathrm{rate}} = (8\,\mathrm{dB},\,2\%) \) using EP strategy. Compared to the conventional baseline, the LightCom framework significantly extends the perceived coverage, with additional gains achieved through importance-aware WF allocation. The coverage gain relative the conventional limit reaches up to $\mathcal{C}_{\mathrm{limit}}^{\mathrm{conv}}$ is $G=0.73$ dB, comprising a power-domain gain of $G_\mathrm{power} = 4.5$ dB and a bandwidth-domain loss of $G_\mathrm{bw} = -3.77$ dB. With the compression rate range of $33\%\sim 50\%$,  corresponding to lossless PNG compression, the coverage gain are approximately $7.5-8$ dB, highlighting strong robustness to low SNR and uncertain SNR scenarios.	Fig.~\ref{fig:robustChannel} further illustrates this robustness advantage, showcasing reconstructed image quality under low SNR conditions at a compression rate of \( r = 9\% \).  These substantial performance improvements
		demonstrate LightCom's exceptional coverage extension, high
		lighting its potential for developing more efficient and resilient
		wireless networks

		\section{Conclusion and Outlook}
	 	 This paper proposed {LightCom}, a novel asymmetric framework for QoE-oriented communications under low SNR conditions. LightCom featured a lightweight transmitter paired with a powerful GenAI-augmented receiver, enabling semantic and perceptual reconstruction beyond bit-level reliability. The design principles of LightCom were systematically analyzed, with a focus on source representation sufficiency and error resilience. A hybrid QoE evaluation methodology was introduced by combining NIQE for perceptual quality and CLIP similarity for semantic coherence, based on which perceived coverage was defined and quantified. To further enhance QoE and extend perceived coverage, we developed importance-aware power allocation strategies for both channel-uncoded and channel-coded scenarios. Simulation results, based on image transmission tasks, validated the effectiveness of LightCom, and demonstrated substantial improvements in robustness and coverage extension compared to traditional QoS-oriented systems. These results underscore the potential of LightCom to advance robust, efficient, and QoE-driven wireless communications.

		These findings point toward several promising directions for future research within the  LightCom framework, including: the development of wireless-tailored GenAI models  to further augment the receiver's capabilities;  optimization of lightweight source and channel coding schems to align with generative reconstruction; optimization of QoE-aware resource allocation strategies to improve  efficiency; and  redesign of QoE-oriented transmission protocol to reduce retransmissions and alleviate network  congestion.

		\bibliographystyle{IEEEtran}
		\bibliography{reference}
		 
	\end{document}